\documentclass[12pt]{article}

\usepackage{epsfig,multicol,multirow}
\usepackage{amsmath,subfigure,latexsym,amssymb}

\def\lsi{\raise0.3ex\hbox{$<$\kern-0.75em\raise-1.1ex\hbox{$\sim$}}}
\def\gsi{\raise0.3ex\hbox{$>$\kern-0.75em\raise-1.1ex\hbox{$\sim$}}}
\def\backder{\raise1.4ex\hbox{$\leftarrow$\kern-0.75em\raise-1.4ex\hbox{$\partial$}}}
\def\forder{\raise1.4ex\hbox{$\rightarrow$\kern-0.75em\raise-1.4ex\hbox{$\partial$}}}

\newcommand{\excleq}{\mathop{\stackrel{!}{=}}}

\newcommand{\lsim}{\mathop{\lsi}}
\newcommand{\gsim}{\mathop{\gsi}}

\newcommand{\be}{\begin{equation}}
\newcommand{\ee}{\end{equation}}
\newcommand{\nn}{\nonumber}
\newcommand{\bea}{\begin{eqnarray}}
\newcommand{\eea}{\end{eqnarray}}
\newcommand{\eps}{\epsilon}
\newcommand{\la}{\langle}
\newcommand{\ra}{\rangle}

\newcommand{\R}{{\kern+.25em\sf{R}\kern-.78em\sf{I} \kern+.78em\kern-.25em}}
\newcommand{\N}{{\kern+.25em\sf{N}\kern-.78em\sf{I} \kern+.78em\kern-.25em}}
\newcommand{\C}{{\kern+.25em\sf{C}\kern-.50em\sf{I} \kern+.50em\kern-.25em}}

\makeatletter
\@addtoreset{equation}{section}
\makeatother

\begin{document}

{
\begin{flushright}
HU-EP-06/12 \\
BI-TP 2006/16 \\
SFB/CPP-06-23
\end{flushright}
}

\vspace*{5mm}

\begin{center}

{\Large{\bf Overlap Hypercube Fermions in QCD}} \\

\vspace*{8mm}

{\Large\bf Simulations Near the Chiral Limit} 

\vspace*{1.5cm}

Wolfgang Bietenholz$^{\rm \, a}$ \ and \ Stanislav Shcheredin$^{\rm \, b}$  \\

\vspace*{8mm}

$^{\rm a}$ Institut f\"{u}r Physik \\
Humboldt Universit\"{a}t zu Berlin \\
Newtonstr.\ 15, D-12489 Berlin, Germany \\

\vspace*{5mm}
$^{\rm b}$ Fakult\"{a}t f\"{u}r Physik \\ 
Universit\"{a}t Bielefeld \\
D-33615 Bielefeld, Germany \\

\end{center}

\vspace*{0.8cm}

The overlap hypercube fermion is a variant of a chirally 
symmetric lattice fermion, which is endowed with a higher level
of locality than the standard
overlap fermion. We apply this formulation in quenched QCD simulations
with light quarks. In the $p$-regime we evaluate the masses of light
pseudoscalar and vector mesons, as well as the pion decay constant
and the renormalisation constant $Z_{A}$. In the $\epsilon$-regime
we present results for the
leading Low Energy Constants of the chiral Lagrangian, $\Sigma$ and
$F_{\pi}$. To this end, we perform fits to predictions by chiral
Random Matrix Theory and by different versions of quenched Chiral
Perturbation Theory, referring to distinct correlation functions.
These results, along with an evaluation of the
topological susceptibility, are also compared to the outcome
based on the standard overlap operator.

\newpage

\section{Overview}

QCD at low energy cannot be handled by perturbation theory.
Therefore it is often replaced by Chiral Perturbation Theory
($\chi$PT) as an effective theory \cite{XPT}. $\chi$PT deals with fields 
for the light mesons (occasionally also nucleons) which are the
relevant degrees of freedom in that regime. Here we present
non-perturbative results for QCD itself --- the fundamental theory --- 
at low energy, and we also establish a connection to the most
important parameters in $\chi$PT. For such systems, chiral symmetry
plays a central r\^{o}le, even though it is only realised approximately.

In general, it is notoriously problematic to regularise quantum field 
theories with massless (or light) fermions in a way, which keeps track of 
the (approximate) chiral
symmetry. For instance, in the framework of dimensional regularisation
this issue is analysed carefully in Ref.\ \cite{Fred}.
However, in the framework of the lattice regularisation there was
substantial progress in this respect at the end of the last century
(for a review, see Ref.\ \cite{SCUJW}).
At least for vector theories a satisfactory solution was found, which
enables the simulation of QCD close to the chiral limit.
But since this formulation is computationally demanding,
production runs are limited to the quenched approximation up to now.

In this work, we present simulation results employing a specific type of
chiral lattice fermions, denoted as the overlap hypercube fermion
(overlap-HF). In Section 2 we briefly review its construction and
discuss some properties, which are superior compared to the standard 
overlap formulation. 

Section 3 is devoted to the regime, where the $p$-expansion of $\chi$PT
\cite{preg} is applicable ($p$-regime). Here we present
simulation results for bare quark masses in the range
from $16.1~{\rm MeV}$ to $161~{\rm MeV}$.
We measure the masses of light pseudoscalar and vector
mesons, and the quark mass according to 
the PCAC relation. The latter fixes the axial-current
renormalisation constant
$Z_{A}$, which has a stable extrapolation to the chiral limit.
We also present a matrix element evaluation of the pion decay
constant $F_{\pi}$, where, however, such an extrapolation is less stable.

In a fixed volume $ V \simeq (1.48 ~ {\rm fm})^{3} \times (2.96 ~ {\rm fm})$,
we further decrease the bare quark mass to $m_{q} \leq 8 ~{\rm MeV}$, 
which takes us into the 
$\epsilon$-regime, i.e.\ the domain of the $\epsilon$-expansion 
\cite{epsreg1} in $\chi$PT. The study of overlap-HFs in 
the $\epsilon$-regime, and also the comparison to the
standard overlap fermions under identical conditions, 
is the main issue of this work, which we present
in Section 4. In that regime, finite size effects and the topological sectors
play an extraordinary r\^{o}le.

Therefore, in Subsection 4.1 we first discuss the distribution of
topological charges --- defined by the fermion index --- and the corresponding
susceptibility. Then we address the Low Energy Constants (LECs) of the
chiral Lagrangian, which parametrise the finite size effects. Hence they
can be extracted even from the $\epsilon$-regime --- i.e.\ from a relatively small
volume --- with their values in infinite volume, which are relevant in physics.
In Subsection 4.2 we compare our results for the leading non-zero eigenvalue
of the Dirac operator to the prediction by chiral Random Matrix Theory
(RMT), which allows for a determination of the chiral condensate
$\Sigma$. 

Finally we return to $F_{\pi}$, which we evaluate in the $\epsilon$-regime
by means of two me\-thods. In Subsection 4.3 we fit the correlation function
of the axial-currents to a prediction by quenched $\chi$PT,
making use of the previously obtained values for $Z_{A}$ and $\Sigma$. 
At last we work directly in the chiral limit (i.e.\ at zero quark mass)
and consider the zero-mode contributions to the pseudoscalar correlation
functions (Subsection 4.4). 
Again, a value for $F_{\pi}$ emerges from a fit of our data to
a prediction by quenched $\chi$PT. However, the two versions of quenched
$\chi$PT that we apply to determine $F_{\pi}$ differ by a subtlety in the 
counting rules for the quenched terms in the $\epsilon$-expansion.

We summarise and discuss our results in Section 5.
Part of them were included previously in a Ph.D.\ thesis
\cite{Stani} and in several proceeding contributions \cite{procs}.

\section{The overlap hypercube fermion in QCD}

The lattice regularisation usually introduces a UV cutoff $\pi /a$,
where $a$ is the lattice spacing. However, a block variable Renormalisation
Group Transformation (RGT) generates a lattice action on a coarser
lattice --- say, with a spacing $a' = 2a$ --- which leaves the system
unchanged, so that still the original cutoff matters \cite{WiKo}. Hence in 
this formulation the lattice artifacts are controlled by $2\pi /a'$, in 
contrast to the situation for a standard action. An infinite iteration of 
this blocking procedure leads in principle (for suitable RGT parameters)
to a {\em perfect lattice action}, which is free of any cutoff artifacts.

In practice, for most systems perfect actions can only be constructed
in some approximation, such as a classical RGT step 
\cite{HasNie,Bern}, and a truncation of the long-range couplings is needed.
Exceptions are, for instance, the quantum rotor \cite{qrot},
the Gross-Neveu model at large $N$
\cite{BFW} and the free fermion, where explicitly parametrised
perfect actions are known \cite{BW96,stagg}. 
Let us start from free Wilson-type 
fermions and iterate the simple (blocking factor $n$) RGT
\bea
e^{-S'[\bar \psi ' , \psi '] } &=& \int {\cal D} \bar \psi {\cal D} \psi \, 
\exp \Big\{ - S[\bar \psi , \psi ] - \frac{\mu}{a} \sum_{x'}
\Big[ \bar \psi'_{x'} - \frac{1}{n^{(d+1)/2}} 
\sum_{x \in x'} \bar \psi_{x} \Big]
\nn \\
& \times & \Big[ \psi'_{x'} - \frac{1}{n^{(d+1)/2}} 
\sum_{x \in x'} \psi_{x}\Big] \Big\} \ .
\label{freeRGT}
\eea
Here $x$ ($x'$) are the sites of the original, fine (blocked, coarse) 
lattice, populated by the spinor fields $\bar \psi, \, \psi$ 
($\bar \psi' , \, \psi '$), with the lattice action $S$ ($S'$),
and $x \in x'$ are the sites $x$ in the $n^{d}$ block with centre $x'$
(in $d$ dimensions, and $n=2,3 \dots$).
$\mu \neq 0$ is a real, dimensionless RGT parameter.
After an infinite number of iterations, a perfect, free lattice action
$S^{*}[ \bar \psi , \psi ]$ emerges, with a Dirac
operator $D$ consisting of a vector term plus a scalar term,
\begin{equation}
S^{*} [ \bar \psi , \psi ] = a^{d} \sum_{x,y} \bar \psi_{x} D_{xy} \psi_{y} \ ,
\quad D_{xy} = \gamma_{\mu} \rho_{\mu} (x-y) + \lambda (x-y) \ ,
\end{equation}
(where we denote the spacing on the final lattice again by $a$).
For a finite RGT parameter $\mu$, the scalar term $\lambda$ is non-zero.
Then this action is {\em local}, 
since the couplings in $\rho_{\mu}$ and $\lambda$
decay exponentially at large distances $| x -y |$
even at zero fermion mass
($x$ and $y$ are lattice sites) \cite{BW96}. This is satisfactory
from the conceptual perspective, but in view of practical applications
the couplings need to be truncated to a short range. A useful truncation 
scheme works as follows \cite{HF}: one first optimises the RGT so that
the locality of the operator $D$ is maximal, hence the truncation is 
minimally harmful. Then one constructs the perfect fermion action on a small,
periodic lattice of size $3^{4}$ (where we now specify $d=4$). 
The couplings obtained in this way are finally used
also in larger volumes. This is a truncation to a {\em hypercube fermion}
(HF) operator $D_{\rm HF}$ with
\begin{equation}
{\rm supp} [ \rho_{\mu}(x-y) ] \ , \
{\rm supp} [ \lambda (x-y) ] \ \subset \ \Big\{ \, x,y \Big| 
| x_{\nu} - y_{\nu} | \leq a \ , \ \forall \nu \, \Big\} \ .
\end{equation}
The free HF still has excellent scaling properties \cite{HF,thermo}.

At mass zero, the perfect fermions are {\em chiral}, since $D$ solves
the Ginsparg-Wilson relation (GWR) \cite{GW}
\begin{equation}  \label{GWR}
D \gamma_{5} + \gamma_{5} D = \frac{a}{\mu} D \gamma_{5} D
\end{equation}
for a simple RGT of the type (\ref{freeRGT}),
with a transformation parameter $\mu \approx 1$.
The GWR is obvious for free perfect fermions, but it also persists under
gauge interaction.
Classically perfect Dirac operators solve the GWR as well 
\cite{HLN}, and the free HF operator $D_{\rm HF}$ does so to 
very good approximation, as the spectrum shows \cite{EPJC}.

In our HF formulation of lattice QCD \cite{HF,Wupp,Stani}
\footnote{For recent applications in thermodynamic systems,
see Ref.\ \cite{Biel}.},
the gauge field is treated rather ad hoc,
since it is the fermionic part which is most delicate. Hence we use the
standard Wilson gauge action, and the Dirac operator is gauged as follows:
if the lattice spinors $\bar \psi_{x}$ and $\psi_{y}$ are coupled by
$D_{{\rm HF}, xy}$ --- 
i.e.\ if they are located in the same lattice unit hypercube ---
we connect them by the shortest lattice paths and multiply the compact 
link variables on these paths. The normalised sum of these link products
is denoted as a {\em hyperlink} between $\bar \psi_{x}$ and $\psi_{y}$
(a fully explicit description is given in Refs.\ \cite{Wupp,Stani}).

In addition, we replace the simple links in the above procedure by
fat links, $U_{\mu}(x) \to (1 - \alpha ) U_{\mu}(x) + \frac{\alpha}{6}
\sum {\rm [staples]} $ , where $\alpha$ is a parameter to be optimised.

This HF is nearly rotation invariant, but its mass
is subject to a significant additive renormalisation \cite{HF}.
Criticality can be approximated again by amplifying all the link variables
by a suitable factor $u > 1$. At last, in the vector term only
they are also multiplied by an extra factor $v \lsim 1$; this hardly alters 
the critical point, but it improves chirality, i.e.\ it moves $D_{\rm HF}$
closer to a GWR solution. The fat link, e.g.\ with $\alpha$ in the range
$0.3 \dots 0.6$, helps further in this respect \cite{China,Stani}.

\begin{figure}[h!]
\centering
\includegraphics[angle=270,width=0.6\linewidth]{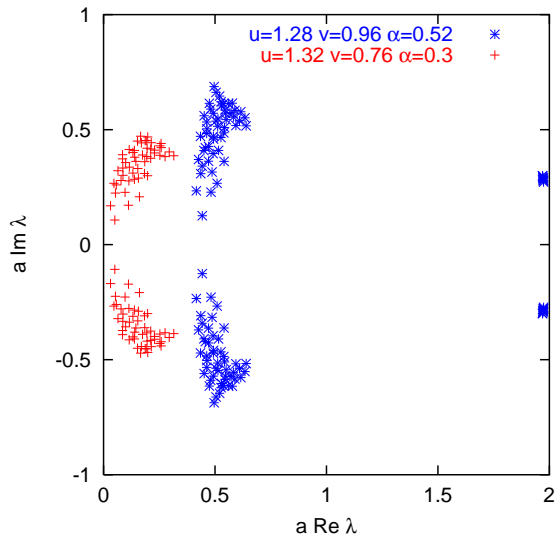}
\caption{{\it The low lying part of the spectrum of $D_{\rm ovHF}$,
plus eigenvalues with maximal real part,
for a typical configuration at $\beta = 5.85$ on a $8^{4}$ lattice.
We show one set of parameters which provides a good approximations 
to criticality and to the spectral Ginsparg-Wilson circle. 
However, for practical purposes we prefer the parameterisation 
(\ref{param}), which leads to the other set of eigenvalues shown
in this Figure.}}
\label{specfig}
\end{figure}

An optimisation of that kind was performed previously for quenched QCD
at $\beta =6$ \cite{ovHF}. Here we present the more difficult construction at
$\beta = 5.85$, which corresponds to a lattice spacing of
$ a \simeq 0.123~{\rm fm}$.
\footnote{Throughout this work, we refer to the Sommer scale \cite{Sommer}
for physical units. We do not keep track of possible errors in that scale.}
As a compromise between various optimisation criteria,
we chose the following parameters \cite{Stani}
\begin{equation}  \label{param}
u = 1.28 \, , \quad v = 0.96 \, , \quad \alpha = 0.52 \, , \quad \mu = 1 \, ,
\end{equation}
and the HF couplings are still those identified for the massless,
truncated perfect, free fermion \cite{HF}.
The physical part of the spectrum 
of this $D_{\rm HF}$ for a typical configuration at $\beta =5.85$
is shown in Fig.\ \ref{specfig}. We also mark some of the eigenvalues 
with maximal real part; they reveal a striking difference from Wilson
fermion spectra, which extend to much larger real parts.
Moreover, we include the low eigenvalues
for an alternative set of parameters, which moves the HF close to
criticality. In that case we recognise a good approximation to
the circle in the complex plane with centre and radius $\mu =1$, 
which characterises an exact solution to the GWR (\ref{GWR}).
However, for the practical purposes in this paper, we prefer
the parameters in eq.\ (\ref{param}) (the advantages are a faster
convergence and a better locality if the HF operator is inserted
in the overlap formula, which we are going to discuss next).

Of course, due to the truncation and the imperfect gauging, the
scaling and chirality  of this HF are somewhat distorted. At least chirality
can be corrected again by inserting $D_{\rm HF}$ into the overlap formula
\cite{Neu}
\begin{equation}  \label{overlap}
D_{\rm ov} = \frac{\mu}{a} \Big( 1 + A / \sqrt{A^{\dagger} A} \Big) \ , 
\qquad A := D_{0} - \frac{\mu}{a} \ .
\end{equation}
This formula yields a solution to the GWR (\ref{GWR}), provided that 
$D_{0}$ is a $\gamma_{5}$-Hermitian lattice Dirac operator, 
$D_{0}^{\dagger} = \gamma_{5} D_{0} \gamma_{5}$. Another condition
is that a value for $\mu$ must be available, which separates 
--- for typical configurations at the gauge coupling under 
consideration  --- the (nearly) real part of the spectrum of
$D_{0}$ in a sensible way into small (physical) eigenvalues and large
eigenvalues (at the cutoff scale). Thus the physical eigenvalues are separated
from the doublers, and we obtain the correct number of flavours.
$\gamma_{5}$-Hermiticity holds for example for the Wilson operator
$D_{\rm W}$ and for $D_{\rm HF}$, and for instance at $\beta =6$ good values
for $\mu$ can easily be found in both cases \cite{HJL,ovHF}.

$D_{0}=D_{\rm W}$ is the kernel of the {\em standard overlap operator}
formulated by H.\ Neuberger \cite{Neu}; we denote it by $D_{\rm N}$. 
In this case, $D_{\rm W}$ changes drastically as it is inserted 
into the overlap formula (\ref{overlap}). 
In contrast, $D_{\rm HF}$ is already an approximate Ginsparg-Wilson
operator, hence the transition to the corresponding overlap operator,
$D_{\rm HF} \to D_{\rm ovHF}$, is only a modest modification,
\begin{equation}  \label{approxGWR}
D_{\rm ovHF} \approx D_{\rm HF} \ ,
\end{equation}
particularly in view of the eigenvalues with large real parts.
In both cases, a lattice modified but exact chirality is guaranteed
\cite{ML98} through the GWR.
The correct axial anomaly is reproduced in all topological sectors
for the standard overlap fermion \cite{DA}, as well as the overlap-HF
\cite{DAWB}.
But other important properties are strongly improved for $D_{\rm ovHF}$
compared to $D_{\rm N}$, due to the construction of $D_{\rm HF}$ and
relation (\ref{approxGWR}). 
\footnote{The concept of improving properties of $D_{\rm ov}$ beyond
chirality by choosing a more favourable kernel than $D_{\rm W}$
was suggested in Ref.\ \cite{EPJC} and meanwhile implemented in
several variants \cite{WBIH,China,plagiat,Bern,filter}.}

Let us first consider the condition number of the Hermitian operator
$A^{\dagger}A$, i.e.\ the argument of the square root in the overlap
formula (\ref{overlap}). In our simulations, we project out the lowest
30 modes of $A^{\dagger}A$ --- which are treated separately --- and we
approximate the remaining part by a Chebyshev polynomial to an absolute
accuracy of $10^{-12}$. The polynomial degree which is needed for a fixed
precision is proportional to the square root of the condition number
\begin{equation}  \label{c31}
c_{31} := \frac{\lambda_{\rm max}}{\lambda_{31}} \ , \quad
{\rm with} \ \lambda_{\rm max} \ \ (\lambda_{31}) \ : \
{\rm maximal ~ (31^{st})~ eigenvalue~of~} A^{\dagger}A \, .
\end{equation}
The values for $\langle c_{31} \rangle$, given in Table 1, show that 
the required polynomial degree for $D_{\rm ovHF}$ is a factor $\approx 5$
lower. The computational effort is roughly proportional to this degree
(the computation of the lowest modes of $A^{\dagger}A$ and their
separate treatment are minor issues in terms of the computing time).
On the other hand, the use of the HF kernel is computationally more 
expensive than $D_{\rm W}$ by about a factor $15$, so that an overhead by a factor
$\approx 3$ remains. We hope for this factor to be more than compensated
by the virtues of $D_{\rm ovHF}$, in particular 
by an improved scaling due to the perfect action background of $D_{\rm HF}$
and relation (\ref{approxGWR}). That property is very well confirmed
for the free fermion \cite{EPJC} and for the 2-flavour Schwinger model
(with quenched configurations, but measurement data re-weighted by the
fermion determinant) \cite{WBIH}. But the corresponding
scaling behaviour in QCD is not explored yet.\\

\begin{table}
\begin{center}
\begin{tabular}{|c|c||c|c|}
\hline
 & $\beta$ & $D_{\rm N}$ & $D_{\rm ovHF}$ \\
\hline
\hline
$\langle c_{11} \rangle$ & $5.85$ & $4266(194)$ & $179(10)$ \\
\hline
$\langle c_{21} \rangle$ & $5.85$ & $1723(46)$ & $73(2)$ \\
\hline
$\langle c_{31} \rangle$ & $5.85$ & $1149(15)$ & $49.2(8)$ \\
\hline
\hline
 $\langle \gamma_{\rm loc} \rangle$ & $5.85$ & $0.63(1)$ 
& $0.73(1)$ \\
\hline
 $\langle \gamma_{\rm loc} \rangle$ & $5.7$ & $0.56(2)$ & 
$0.76(5)$ \\
\hline
 $\langle \gamma_{\rm loc} \rangle$ & $5.6$ & --- & $0.53(3)$ \\
\hline
\end{tabular}
\end{center}
\caption{\it The characteristic indicators for
the kernel condition number and for the locality
of the overlap operators $D_{\rm N}$ (standard) and $D_{\rm ovHF}$ 
(described in the text), on a $12^{3} \times 24$ lattice.}
\label{tabrotcloc}
\end{table}

\begin{figure}[h!]
\centering
\hspace*{3mm} \includegraphics[angle=270,width=0.65\linewidth]{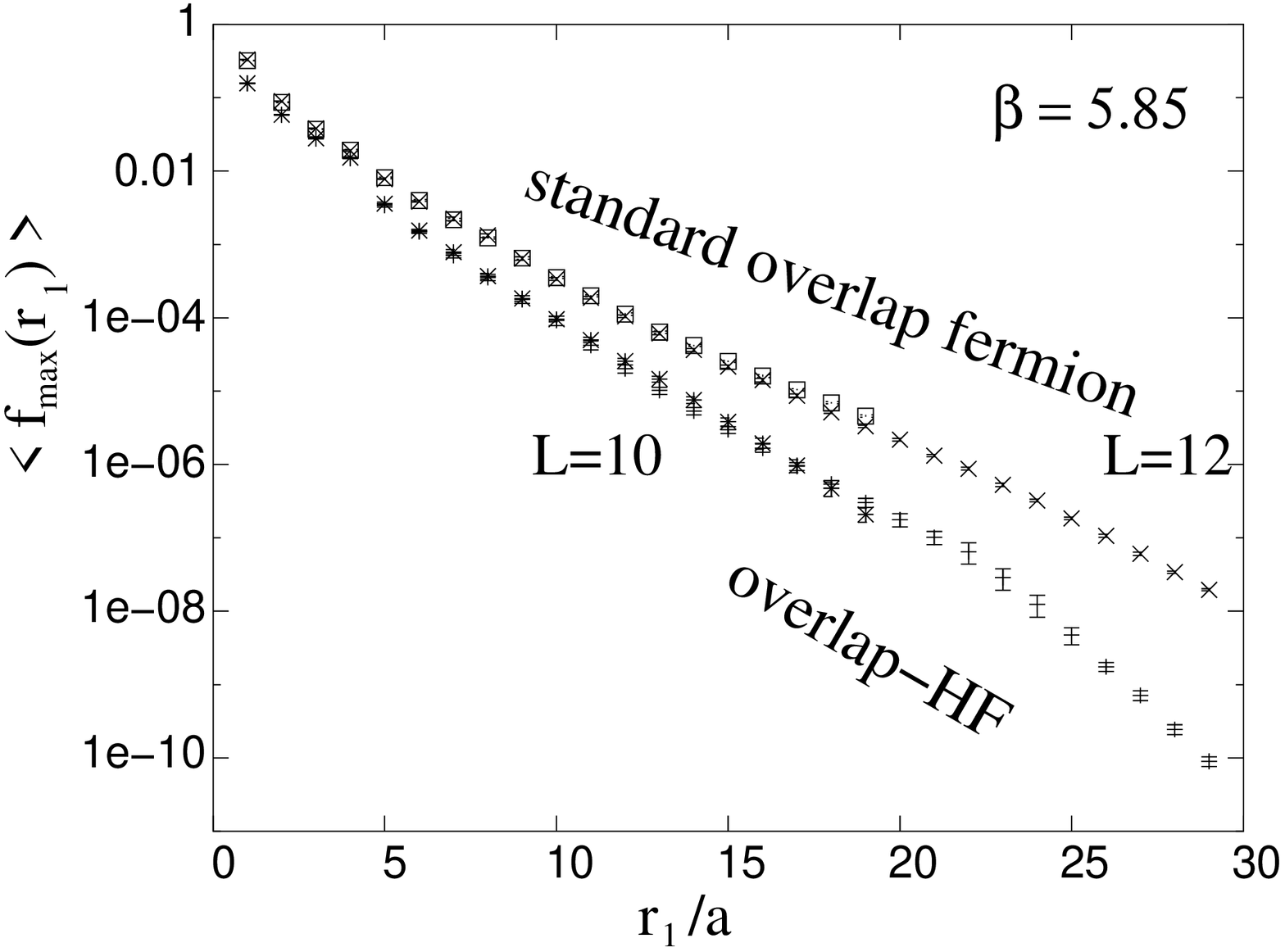} \\
\includegraphics[angle=270,width=0.75\linewidth]{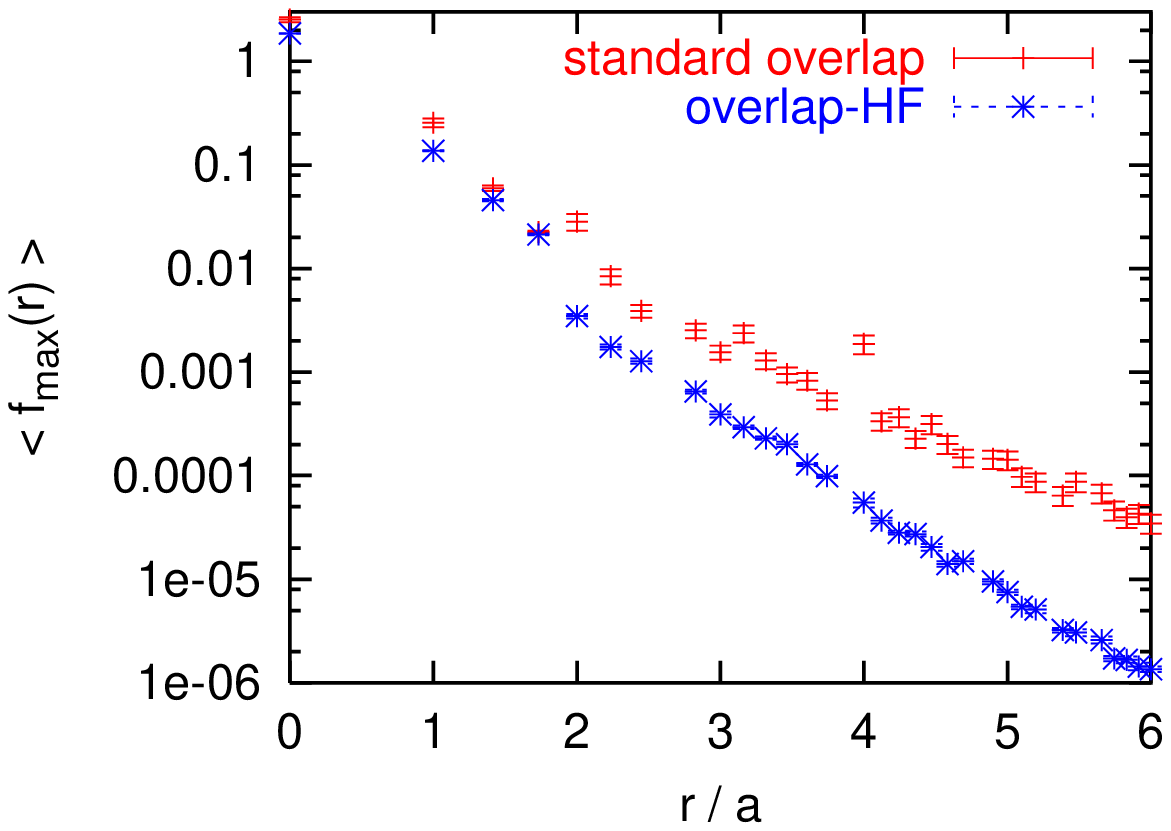}
\caption{{\it The locality of different overlap fermions, measured by the 
maximal impact of a unit source $\eta_{x}$ over 
a taxi driver distance $r_{1}$ 
(on top), and over a Euclidean distance $r$ 
(below), at $\beta =5.85$. We see that the overlap-HF
is clearly more local. The upper plot also shows
that the slope hardly changes if we proceed from
a $10^{4}$ to a $12^{3} \times 24$ lattice. 
(The increasing slope beyond $r_{1} =18$ is a consequence
of the anisotropy of the $12^{3} \times 24$ lattice.)
The lower plot illustrates additionally that the overlap-HF decay follows closely
an exponential in the Euclidean distance, which indicates a good approximation
to rotation symmetry.}}
\label{localfig1}
\end{figure}

An important aspect that we do explore here is the locality.
For this purpose, we put a unit source at the origin,
$\eta_{x} = \delta_{x,0}$, and measure the expectation value of the
function \cite{HJL}
\be
f_{\rm max}(r_{1}) := \ ^{\rm max}_{\ \, y} 
\left. \Big\{ \Vert D_{{\rm ov},yx}(U) \eta_{x} \Vert \ \right| 
\ \Vert y \Vert_{1} = r_{1} \Big\} \ .
\label{fmax}
\ee
Following the usual convention
we take here the distance in the taxi driver metrics 
$\Vert y \Vert_{1} := \sum_{\mu} | y_{\mu} |$.
A comparison, still at $\beta =5.85$, is given in Fig.\ \ref{localfig1} 
(on top) and Table \ref{tabrotcloc};
we see a clear gain for $D_{\rm ovHF}$, 
i.e.\ a decay of $f (r_{1} ) \propto \exp ( - \gamma_{\rm loc} r_{1} )$
(beyond short $r_{1}$) where $\gamma_{\rm loc}$ is larger
for $D_{\rm ovHF}$ than for $D_{\rm N}$.
If we refer to the Euclidean distance $r = | y |$, the maximum
is taken over different sets, but the resulting decay is still
exponential. In the lower plot of Fig.\ \ref{localfig1}
we see that the decay for $D_{\rm ovHF}$ is not only faster, but it also
follows in a smoother way the exponential shape in $r$, which hints at
a good approximation to rotation symmetry.

Locality is a vital requirement
for a safe continuum limit in lattice field theory \cite{HJL}.
This poses a limitation on the overlap operator (\ref{overlap}), since
for decreasing $\beta$, at some point an exponential decay is not
visible any longer.
\footnote{The question, in which sense locality still persists in a conceptual
sense is discussed in Ref.\ \cite{Shamir}. In any case, if no exponential
decay is clearly visible, the operator should certainly not be applied in
simulations.}

\begin{figure}[h!]
\centering
\includegraphics[angle=270,width=0.8\linewidth]{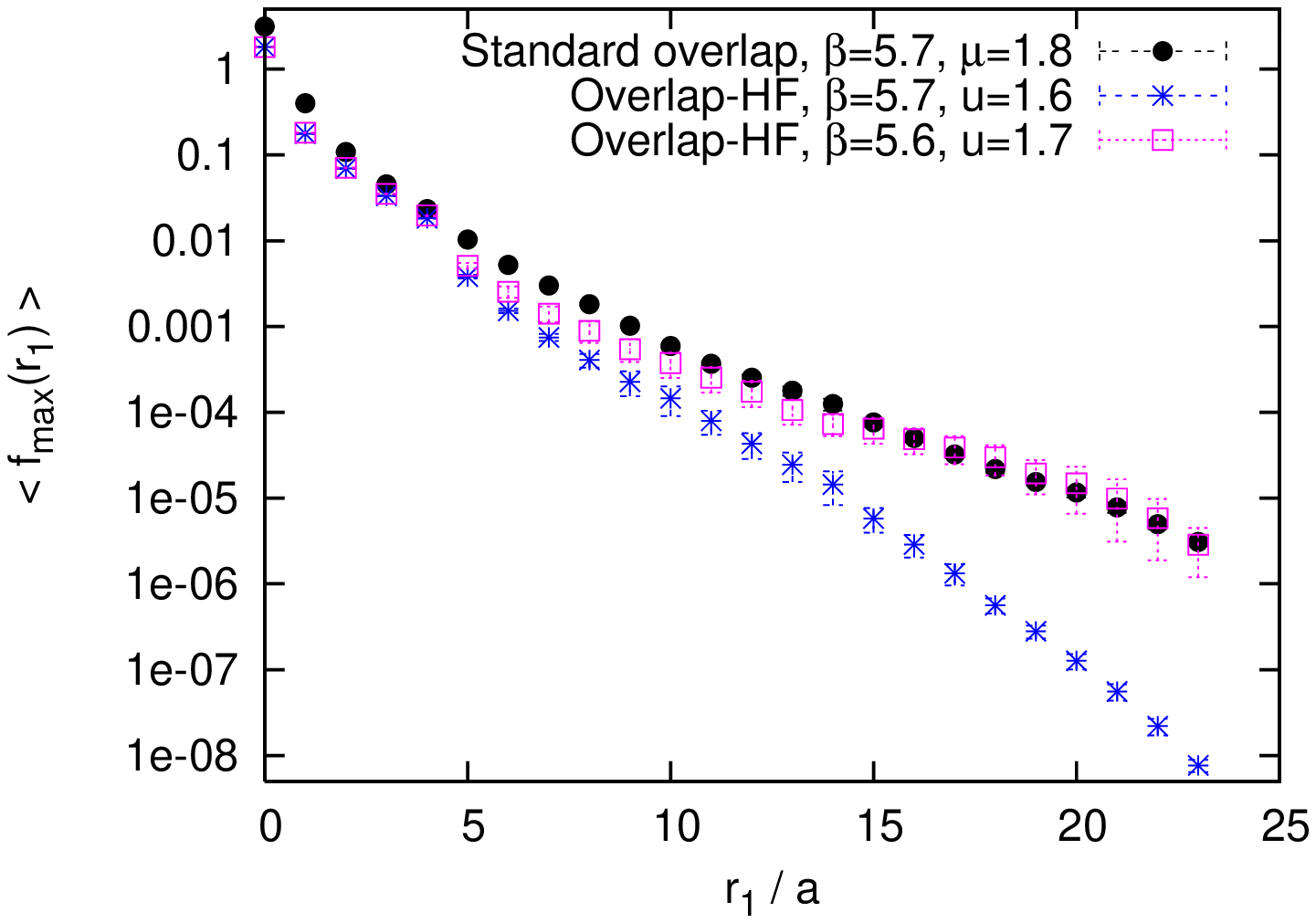}
\caption{{\it The locality of different overlap fermions, measured by the 
maximal impact of a unit source $\eta_{x}$ over a taxi
driver distance $r_{1}$, 
on a $12^{3} \times 24$ lattice. 
For the overlap-HF we see a superior locality
at $\beta = 5.7$. Its locality persist up to $\beta =5.6$,
where no exponential decay is visible for the standard overlap formulation.}}
\label{localfig2}
\end{figure}

Fig.\ \ref{localfig2} shows the situation at $\beta=5.7$ 
(i.e. $ a \simeq 0.17 ~ {\rm fm}$), where
both operators are still local (with optimal parameters
$\mu$ and $u$ for $D_{\rm N}$ resp.\ $D_{\rm ovHF}$),
but again the locality of $D_{\rm ovHF}$ is superior. As we push further
to $\beta = 5.6$, we do not observe an exponential decay for $D_{\rm N}$ anymore,
but $D_{\rm ovHF}$ is still manifestly
local (even without adjusting the parameters
in eq.\ (\ref{param}), except for $u$).
Therefore, the overlap-HF provides {\em chiral fermions on coarser 
lattices.}

\section{Applications in the $p$-regime}

To handle strong interactions at low energy, 
$\chi$PT replaces QCD by an effective Lagrangian. It involves
the terms allowed by the symmetries, which are ordered
according to a low energy hierarchy.
To its leading order, the effective Lagrangian of $\chi$PT
takes the form
\be  \label{Leff}
{\cal L}_{\rm eff}[U] = \frac{F_{\pi}^{2}}{4} \, {\rm Tr} \,
\Big[ \partial_{\mu} U(x)^{\dagger} \partial_{\mu} U (x) \Big]
- \frac{\Sigma m_{q}}{2} \, {\rm Tr} \, \Big[ U(x) + U(x)^{\dagger} \Big] 
+ \dots \ . 
\ee
Throughout this work, we assume a degenerate quark mass $m_{q}$ for
all flavours involved.
The LECs are the coefficients attached to the
terms in this expansion; in the leading order $F_{\pi}$ and $\Sigma$.
The effective theory does have its predictive power, but
the values of the LECs cannot be determined within $\chi$PT.
To this end, one has to return to QCD as the fundamental theory,
and it is our main goal in this work to investigate how far the LECs 
can be obtained from lattice QCD simulations.

In $\chi$PT, one usually considers a finite spatial box,
which has 
a volume $L^{3}$, 
and one expects the meson momenta $p$ to be small, so that
\be  \label{4piFpi}
p \sim \frac{2\pi}{L} \ll 4 \pi F_{\pi} \ .
\ee
Note that the term $4 \pi F_{\pi}$ takes
a r\^{o}le analogous to $\Lambda_{\rm QCD}$.

$\chi$PT then deals with a perturbative scheme for 
the momenta and masses of the light mesons, with appropriate counting 
rules. The most wide-spread variant of $\chi$PT assumes the volume to be
large, $ L \gg \xi = 1 / m_{\pi}$ (where $\xi$ is
the correlation length, given by the inverse pion mass). Then the 
{\em $p$-expansion} \cite{preg} can be applied.
This is an expansion in the following dimensionless ratios,
which are expected to be small and counted in the same order,
\be
\frac{1}{L F_{\pi}} \sim \frac{p}{\lambda_{\rm QCD}} \sim
\frac{m_{\pi}}{\lambda_{\rm QCD}} \ .
\ee

In this Section we present our simulation results  in the $p$-regime.
We apply the overlap-HF described in Section 2, at $\beta = 5.85$ on a
lattice of size $12^{3} \times 24$, which corresponds to a physical 
volume of $V \simeq (1.48 ~{\rm fm})^{3} \times 2.96~{\rm fm}$.
We evaluated 100 propagators for each of the bare quark masses
$$
a m_q = 0.01, \ 0.02, \ 0.04, \ 0.06, \ 0.08 \quad {\rm and} \quad 0.1 \ 
$$
(in physical units: $16.1~{\rm MeV} \dots 161~{\rm MeV}$), using
a Multiple Mass Solver.
We will see that the smallest mass in this set is at the
edge of the $p$-regime --- even smaller quark masses will be considered
in Section 4. 

We include $m_{q}$ in the overlap operator (\ref{overlap}) in the 
usual way,
\begin{equation}
D_{\rm ov}(m_{q}) = \Big( 1 - \frac{a m_{q}}{2 \mu} \Big) D_{\rm ov} + m_{q} \ ,
\end{equation}
which leaves the largest real overlap Dirac eigenvalue 
invariant.
$m_{q}$ represents the bare mass for the quark flavours $u$ and $d$.

We first evaluate the pion mass in three different ways:

\begin{itemize}

\item $m_{\pi,PP}$ is obtained from the decay of the pseudoscalar correlation
function $\langle P(x) P(0) \rangle$, with 
$P(x) = \bar \psi_{x} \gamma_{5} \psi_{x}$.
This is the most obvious method, but it is not the best one in this case,
as we will see.

\item $m_{\pi,AA}$ is extracted from the decay of the axial-vector correlation
function $\langle A_{4}(x) A_{4}(0) \rangle$, with 
$A_{4}(x) = \bar \psi_{x} \gamma_{5} \gamma_{4} \psi_{x}$.

\item $m_{\pi,PP-SS}$ is obtained from the decay of the difference
$\langle P(x) P(0) - S(x) S(0) \rangle$, where
$S(x) = \bar \psi_{x} \psi_{x}$ is the scalar density.
This subtraction is useful especially at small $m_{q}$, where configurations
with zero-modes ought to be strongly suppressed by the fermion determinant.
In our quenched study, this
suppression does not occur as it should, but the above subtraction in the
observable eliminates the zero-mode contributions, which are mostly unphysical.

\end{itemize}

\begin{figure}[h!]
  \centering
  \includegraphics[angle=270,width=0.9\linewidth]{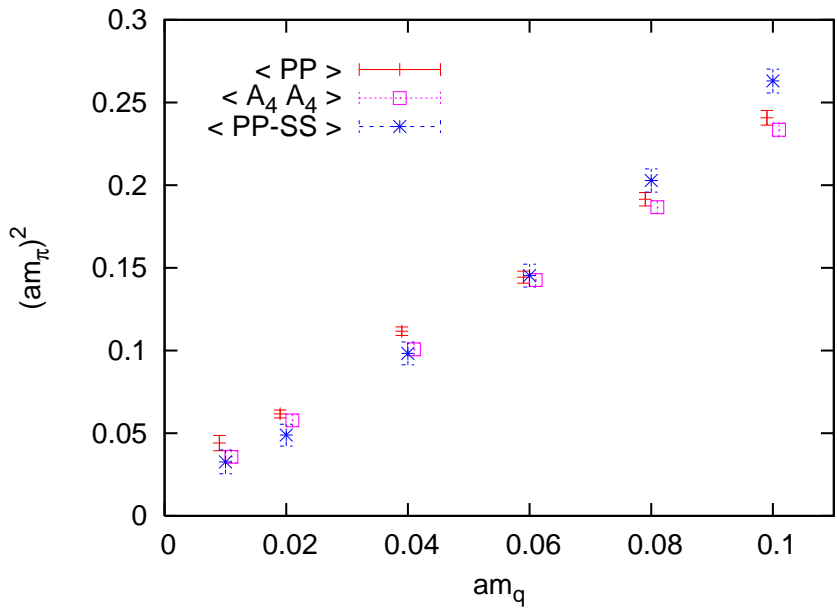}
\caption{{\it The pion mass evaluated from overlap HFs in the
$p$-regime in three different ways, as described in Section 3.}}
\label{pimass}
\end{figure}

We present the results in Fig.\ \ref{pimass}
and Table \ref{tabpreg} (the latter collects
all the results of this Section). The pion masses follow to a good 
approximation the expected behaviour $m_{\pi}^{2} \propto m_{q}$.
Deviations show up at the smallest masses, where we observe the hierarchy
\be
m_{\pi,PP} > m_{\pi,AA} > m_{\pi,PP-SS} \ ,
\ee
in agreement with 
Ref.\ \cite{Bern}.
This shows that the scalar density subtraction is in fact profitable,
since it suppresses the distortion of the linear behaviour down
to the lightest pion mass in Fig.\ \ref{pimass}, 
\be
m_{\pi ,PP-SS} (a m_{q}=0.01) = (289 \pm 32) ~ {\rm MeV} \ .
\ee
That mass corresponds to a ratio $ L / \xi \approx 2$, hence around
this point we are indeed leaving the $p$-regime. 
Based on the moderate quark masses in Fig.\ \ref{pimass}, we find an
impressively small intercept in the
chiral extrapolation,
\be
m_{\pi , PP-SS} (m_{q} \to 0) = (-2 \pm 24) ~ {\rm MeV} \ .
\ee
On the other hand, at our larger $m_{q}$ values the
hierarchy changes due to the interference of the scalar correlator. 

\begin{table}
\hspace*{-0.9cm}
\begin{tabular}{|c||c|c|c|c|c|c|}
\hline
$a m_{q} $ & $0.1$ & $0.08$ & $0.06$ & $0.04$ & $0.02$ & $0.01$ \\
\hline
\hline
$a m_{\pi,PP}$ & $0.491(5)$ & $0.438(5)$ & $0.380(5)$ & $0.314(4)$ 
& $0.248(5)$ & $0.210(11)$ \\
\hline
$a m_{\pi,AA}$ & $0.483(4)$ & $0.432(5)$ & $0.378(5)$ & $0.317(6)$ 
& $0.240(7)$ & $0.189(9)$ \\
\hline
$a m_{\pi,PP-SS}$ & $0.513(7)$ & $0.453(8)$ & $0.381(9)$ & $0.313(11)$ 
& $0.221(15)$ & $0.181(20)$ \\
\hline
\hline
$a m_{\rho}$ & $0.745(7)$ & $0.717(8)$ & $0.694(10)$ & $0.677(15)$ 
& $0.674(32)$ & $0.672(55)$ \\
\hline
$a m_{\rm PCAC}$ & $0.089(3)$ & $0.070(3)$ & $0.052(3)$ & $0.035(3)$ 
& $0.017(2)$ & $0.0087(13)$ \\
\hline
$Z_{A}$ & $1.13(4)$ & $1.14(5)$ & $1.15(6)$ & $1.16(9)$ & $1.16(14)$
& $1.14(18)$ \\
\hline
$a F_{\pi,PP}$ & $0.117(2)$ & $0.114(3)$ & $0.110(3)$ & $0.105(4)$ & $0.103(4)$
& $0.098(4)$ \\
\hline
$a F_{\pi,PP-SS}$ & $0.120(2)$ & $0.114(3)$ & $0.109(3)$ & $0.104(5)$ & $0.010(10)$
& $0.082(12)$  \\
\hline
\end{tabular}
\caption{\it Our results in the $p$-regime for the pion mass, the $\rho$-meson 
mass, the PCAC quark mass, the renormalisation constant $Z_{A}$ and the pion decay
constant $F_{\pi}$, as described in the text and plotted in the Figures
of Section 3 (Figs. \ref{pimass} $\dots$ \ref{Fpifig}).
These results are obtained from 100 propagators at each of the
bare quark masses $m_{q}$.
All the dimensional numbers in this Table are given in 
lattice units at \ $a \simeq 0.123 ~{\rm fm} = (1610 ~ {\rm MeV})^{-1}$.}
\label{tabpreg}
\end{table}

Due to quenching, one expects at small quark masses a logarithmic
behaviour of the form
\begin{equation}  \label{chilog}
\frac{a m_{\pi}^{2}}{m_{q}} = C_{1} + C_{2} \ln a m_{q} + C_{3} a m_{q} \ ,
\quad (C_{1},C_{2},C_{3} :~ {\rm constants}) \ .
\end{equation}
\begin{figure}[h!]
  \centering
  \includegraphics[angle=270,width=.9\linewidth]{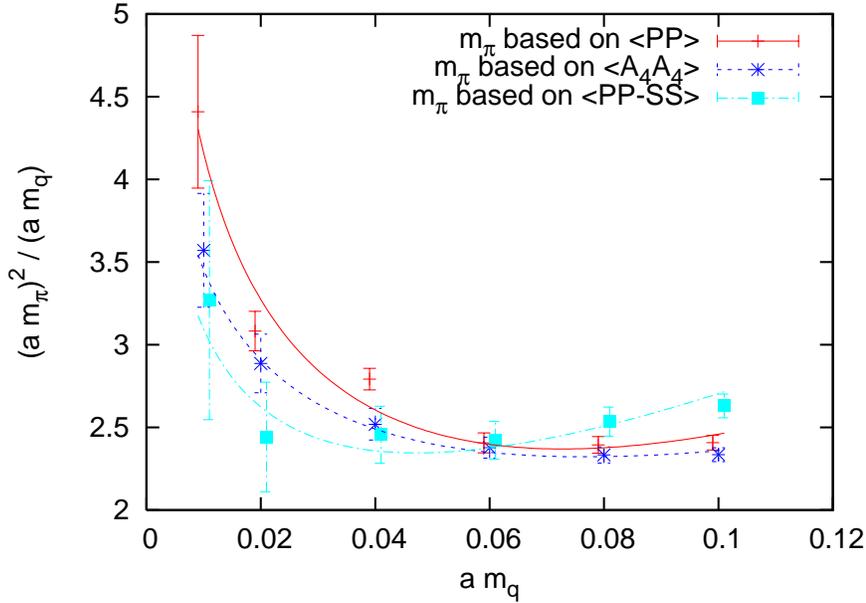}
 \caption{{\it The pion masses of Fig.\ \ref{pimass}, fitted against
the form (\ref{chilog}), which is expected 
for the logarithmic quenching artifacts
in the absence of an additive mass renormalisation.}}
\label{chilogfit}
\end{figure}
Corresponding results are given for instance in Refs.\ 
\cite{Bern2,SchierLat05}. Fig.\ \ref{chilog} shows the fits of our data
to eq.\ (\ref{chilog}). $m_{\pi,AA}$ is best compatible 
with this rule (this property was also hinted at in
Ref.\ \cite{SchierLat05}). At least the deviation 
for $m_{\pi,PP-SS}$ (although inside the errors)
could be expected, since the scalar subtraction
alleviates the quenching artifacts, which give rise to
the logarithmic corrections according to formula (\ref{chilog}).\\

\begin{figure}[h!]
  \centering
  \includegraphics[angle=270,width=0.9\linewidth]{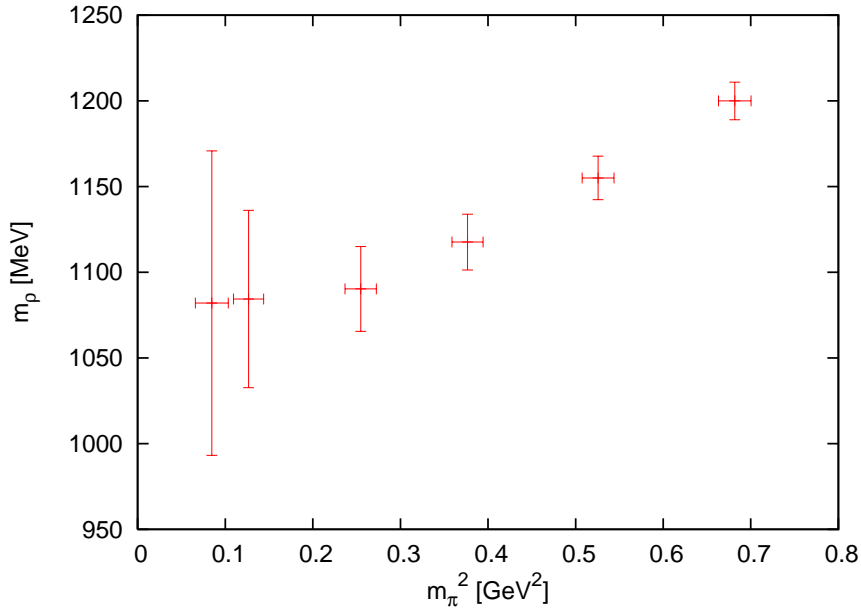}
 \caption{{\it The mass of the $\rho$-meson against the squared
pion mass $m_{\pi, PP-SS}^{2}$.}}
\label{rhomass}
\end{figure}

In Fig.\ \ref{rhomass} (and Table \ref{tabpreg})
we consider the vector meson mass $m_{\rho}$ based
on the same 100 configurations. A chiral extrapolation leads to 
\be
m_{\rho} = (1017 \pm 40) ~ {\rm MeV} \ .
\ee
This agrees well with a
study using the standard overlap operator on the same lattice
\cite{XLF}. It is quite large, however, not only in view of phenomenology,
but also compared to other quenched results in the literature,
which were summarised in Ref.\ \cite{Arifa}. 
\footnote{Note that the difference from the data by other groups ---
obtained with various (mostly non-chiral)
lattice formulations --- is approximately constant 
in $m_{\pi}$ over the range shown here, hence finite size effects
can hardly be blamed for it.}

On the other hand, if we insert our measured ratio 
$m_{\rm pseudoscalar} / m_{\rm vector}$ at $m_{q}=0.01$
into the phenomenologically motivated interpolation formula
presented in Ref.\ \cite{ELae}, then we arrive at a significantly
lower value of $m_{\rho} \approx 789~{\rm MeV}$.

\begin{figure}[h!]
  \centering
  \includegraphics[angle=270,width=.9\linewidth]{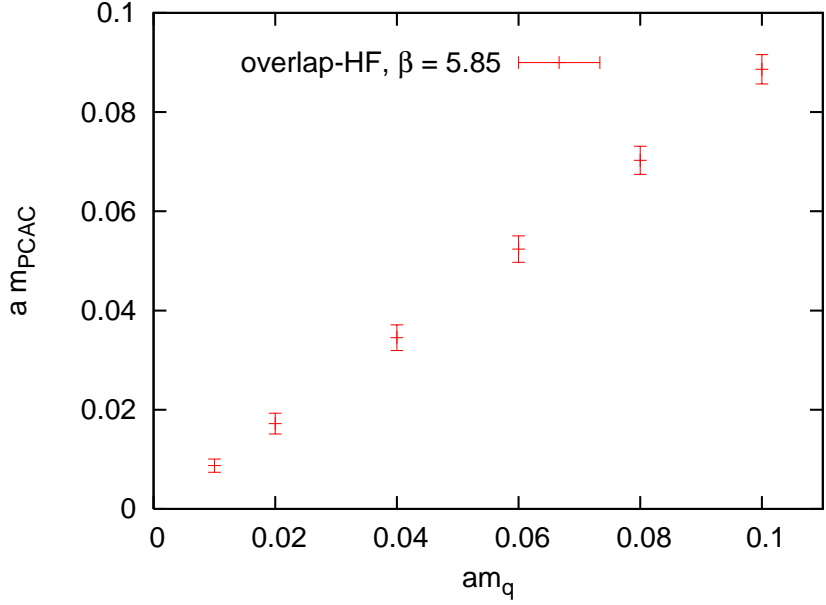}
 \caption{{\it The PCAC quark mass, evaluated for $D_{\rm ovHF}$
according to eq.\ (\ref{PCAC}), from stable plateaux in the time direction.
We find $m_{\rm PCAC}$ values close to $m_{q}$, in contrast to the results 
with the standard overlap operator.}} 
\label{mPCAC}
\end{figure}

\begin{figure}[h!]
  \centering
  \includegraphics[angle=270,width=.9\linewidth]{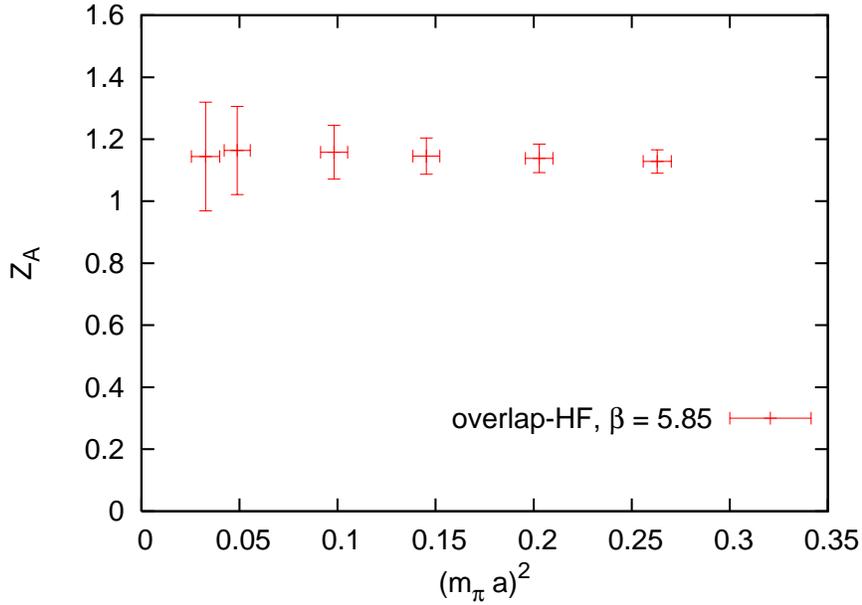}
 \caption{{\it The axial-current renormalisation constant 
$Z_{A}$, determined from eq.\ (\ref{ZAeq}), 
which is found to be close 
to $1$ for the overlap-HF. $Z_{A}$ is plotted
against $m_{\pi , PP-SS}^{2}$.}}
\label{ZAfig}
\end{figure}

Fig.\ \ref{mPCAC} shows the quark mass obtained from the PCAC relation,
\begin{equation}  \label{PCAC}
m_{\rm PCAC} = \frac{ \sum_{\vec x} \langle
(\partial_{4} A_{4}^{\dagger} (x)) P(0) \rangle }
{ \sum_{\vec x} \langle P^{\dagger} (x) P(0) \rangle } \ ,
\end{equation}
which follows closely the bare quark mass $m_{q}$
(we use a symmetric nearest-neighbour difference for $\partial_{4}$).
As a consequence, the axial-current renormalisation constant
\begin{equation}  \label{ZAeq}
Z_{A} = \frac{m_{q}}{m_{\rm PCAC}}
\end{equation}
is close to $1$, see Fig.\ \ref{ZAfig} and Table \ref{tabpreg}. 
A chiral extrapolation leads smoothly to
\begin{equation}
Z_{A} = 1.17(2) \ ,
\end{equation}
which is in contrast to the unpleasantly large values found for the
standard overlap operator: e.g.\ at the same $\beta = 5.85$
and $\mu =1.6$ (a preferred value in that case) it amounts to
$Z_{A} \simeq 1.45$ \cite{XLF,JapZA,Babich}, and (somewhat surprisingly)
at $\beta =6, \ \mu =1.4$ it even rises to $Z_{A} \simeq 1.55$
\cite{Babich}. According to Ref.\ \cite{filter}, the fat link
could be especially helpful for the property $Z_{A} \approx 1$, which 
is favourable for a connection to perturbation theory.

\begin{figure}[h!]
  \centering
\includegraphics[angle=270,width=0.9\linewidth]{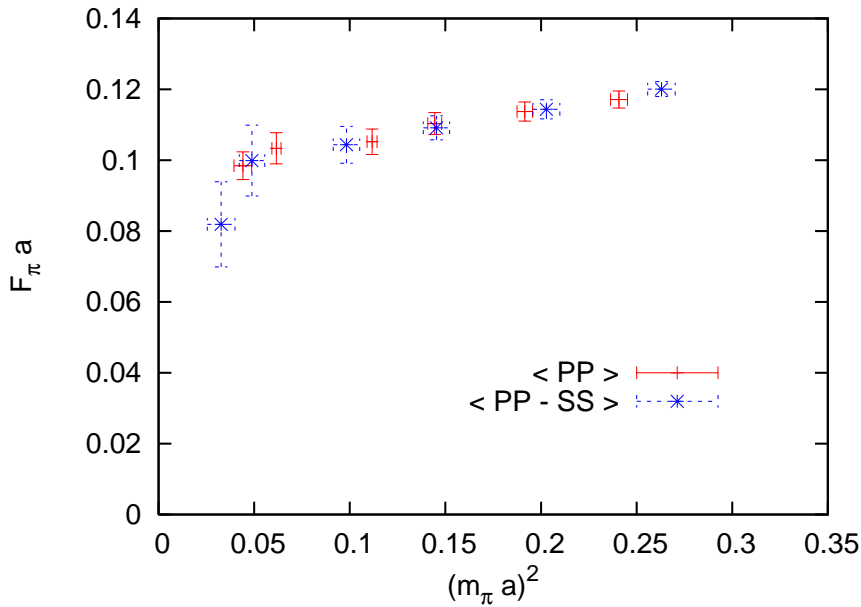}
\caption{{\it The pion decay constant based on a matrix element evaluation 
in the $p$-regime --- given by eq.\ (\ref{Fpip}) --- using the overlap-HF.}}
\label{Fpifig}
\end{figure}

As a last observable in the $p$-regime, we measured the pion decay constant
by means of the (indirect) relation
\begin{equation}  \label{Fpip}
F_{\pi} = \frac{2 m_{q}}{m_{\pi}^{2}} \ \left| \langle 0 |
P | \pi \rangle \right| \ ,
\end{equation}
based on $P(x)P(0)$, or based on $P(x)P(0) - S(x)S(0)$
(in eq.\ (\ref{Fpip}) this affects both, the denominator and the
pion state). 
\footnote{A recent alternative attempt to evaluate LECs from simulations
in the $p$-regime was presented in Ref.\ \cite{notchiral}.}
The results are given in Fig.\ \ref{Fpifig}
and Table \ref{tabpreg}. In particular the value at 
$am_{q}=0.01$ (the lightest quark mass in this plot)
is significantly lower for the case of the scalar 
subtraction. Hence this pushes the result towards the chirally
extrapolated phenomenological value of $\approx 86 ~ 
{\rm MeV}$ \cite{CoDu}.

However, a chiral extrapolation based on these data for $F_{\pi}$
would be risky ---
apparently we are too far from the chiral limit for this purpose.
An extrapolated value of $F_{\pi,PP}$  would come out clearly too large,
as it is also the case for $D_{\rm N}$ \cite{XLF}.
But in particular the instability of $F_{\pi ,PP-SS}$ 
at $a m_{q} = 0.04, \ 0.02, \ 0.01$ calls for a clarification by
yet smaller quark masses. We did consider still much smaller values of $m_{q}$
in the same volume. As the results for the pion masses suggest, we are thus
leaving the $p$-regime. For the tiny masses $a m_{q} \leq 0.005$
we enter in fact the $\epsilon$-regime, where observables like $F_{\pi}$ 
have to be evaluated in a completely different manner. This is the
subject of Section 4.

\section{Applications in the $\epsilon$-regime}

The $\epsilon$-regime of QCD is characterised by a relatively
small volume, i.e.\ the correlation length $\xi$ exceeds the linear
box size $L$. On the other hand, the box still has to be large 
compared to the scale given by $4 \pi F_{\pi}$, as in eq.\ (\ref{4piFpi}),
so that the partition
function is saturated by the lowest states only --- higher states
(above about $1 ~ {\rm GeV}$) do not contribute significantly.
This amounts to the condition
\be
\frac{1}{m_{\pi}} > L \gg \frac{1}{4\pi F_{\pi}} \ .
\ee
In such a box, the $p$-expansion of
$\chi$PT fails, in particular due to the dominant r\^{o}le of the
zero-modes. However, the latter can be treated separately by means of
collective variables, and the higher modes --- along with the pion mass ---
are then captured by the {\em $\epsilon$-expansion} \cite{epsreg1}.
One now counts the ratios
\be  \label{epsexp}
\frac{m_{\pi}}{\Lambda_{\rm QCD}} \sim \frac{p^{2}}{\Lambda_{\rm QCD}^{2}}
\sim \frac{1}{( L F_{\pi})^{2}}
\ee
as small quantities in the same order.

The Haar measure for the fields $U(x) \in SU(N_{f})$
in the coset space of the chiral symmetry breaking,
\be
SU(N_{f})_{L} \otimes SU(N_{f})_{R} \to SU(N_{f})_{L+R} \qquad
({\rm for}~N_{f}~{\rm flavours}) \ ,
\ee
is worked out in Ref.\ \cite{epsmeas}. The corresponding
non-linear $O(N)$ $\sigma$-model (with a symmetry breaking to
$O(N-1)$) in a small box was studied with the Faddeev-Popov method
to two loops \cite{epsFP}, and with the Polyakov functional measure to three 
loops \cite{epsPoly}.

This setting cannot be considered a physical situation.
Nevertheless there is a strong motivation for its numerical study:
the point is that the finite size effects are parametrised by the
LECs of the effective chiral Lagrangian as they occur in infinite volume, hence
the physical values of the LECs can (in principle) be evaluated even in an
unphysically small box.

We recall that the effective chiral Lagrangian includes all terms which
are compatible with the symmetries, ordered according to suitable low
energy counting rules, in this case the counting
rules (\ref{epsexp}) of the $\epsilon$-expansion. 
We further presented in Section 3
the LECs as coefficients of these
terms, for instance $F_{\pi}$ and $\Sigma$ in the leading order.
Their determination from QCD --- the fundamental
theory --- at low energy is a challenge for lattice simulations. 
As a test, such a lattice determination in the 
$\epsilon$-regime has been performed successfully for the
$O(4)$ $\sigma$-model (which describes chiral symmetry breaking with
two flavours) some years ago \cite{epsO4}, but
in QCD this method \cite{GHLW} had to await the advent of chiral lattice
fermions. Unfortunately, the quenched results for the LECs are
affected by (mostly logarithmic) finite size effects \cite{loga}, 
so that the final results
by this method still have to wait for the feasibility of QCD simulations
with dynamical, chiral quarks.

A peculiarity of the $\epsilon$-regime is that the topological
sector plays an essential r\^{o}le \cite{LeuSmi}: if one measures
observables in a specific sector, the expectation values often
depend significantly on the (absolute value of the) topological
charge in this sector. In particular for the evaluation of the LECs,
a numerical measurement inside a specific sector and a 
confrontation with the analytical predictions in this sector is in 
principle sufficient. This requires the collection of a large number of
configurations in a specific sector. The ``topology
conserving gauge actions'' \cite{topogauge} are designed to facilitate 
this task. However, here we stay with the Wilson gauge action, which allows
us to investigate also the statistical distribution of the
topological charges, which we are going to address next.

\subsection{The distribution of topological charges}

A priori, it is not obvious how to introduce topological sectors
in the set of lattice gauge configurations. However, if one deals with
Ginsparg-Wilson fermions, a sound definition is given by adapting
the Atiyah-Singer Theorem from the continuum and defining
the topological charge by the fermionic index $\nu$ \cite{HLN},
\be
{\rm topological~charge} \, \excleq \, \nu := n_{+} - n_{-} \ ,
\ee
where $n_{\pm}$ is the number of zero-modes with positive/negative
chirality. We remark that these numbers are unambiguously determined once
a Ginsparg-Wilson Dirac operator is fixed (and that in practice
only chirality positive or chirality negative zero-modes occur
in one configuration~\footnote{This property even holds 
in cases where ``cooling'' deforms the configuration
into a form, where a semi-classical picture suggests the presence of
topological objects with opposite chiralities \cite{KINLat05}.
Regarding the fermion index, a cancellation happens for instance for
free Ginsparg-Wilson fermions, but in a realistic gauge background
such an unstable constellation is very unlikely.}
). However, for a given gauge configuration, the 
index for different Ginsparg-Wilson operators
does not need to agree. Albeit the level of agreement should be high
for smooth configurations, i.e.\ it should 
--- and it does 
\footnote{For instance, we observed at $\beta = 6.15$ on a $16^{4}$
lattice that the index of $D_{\rm N}$ is very stable as $\mu$ rises
from $1.3$ to $1.7$; this changes less than $2 \%$ of the indices.}
---
increase for rising values 
of $\beta$. 

\begin{figure}[h!]
  \centering
\includegraphics[angle=270,width=.9\linewidth]{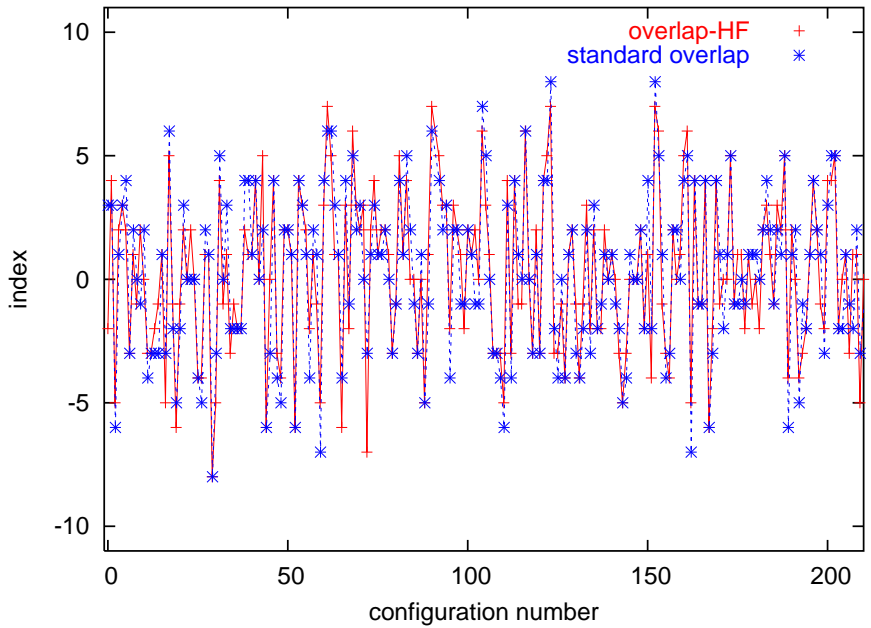}
\caption{{\it Index histories for $D_{\rm ovHF}$ (see Section 2)
and for $D_{\rm N}$ (at $\mu =1.6$) for the same set of configurations.}}
\label{tophistory}
\end{figure}

\begin{figure}[h!]
  \centering
\includegraphics[angle=270,width=.5\linewidth]{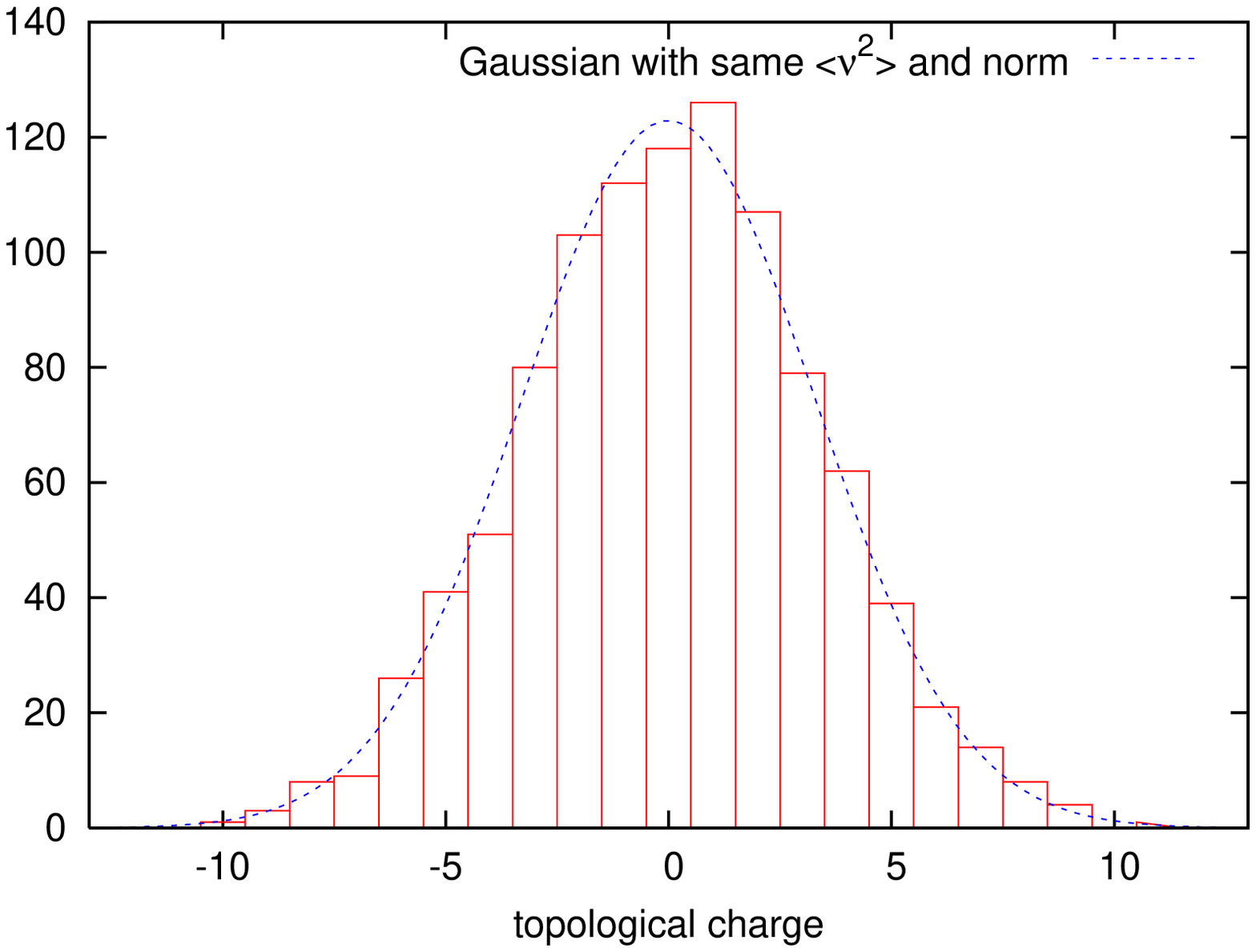}
\hspace*{-5mm}
\includegraphics[angle=270,width=.5\linewidth]{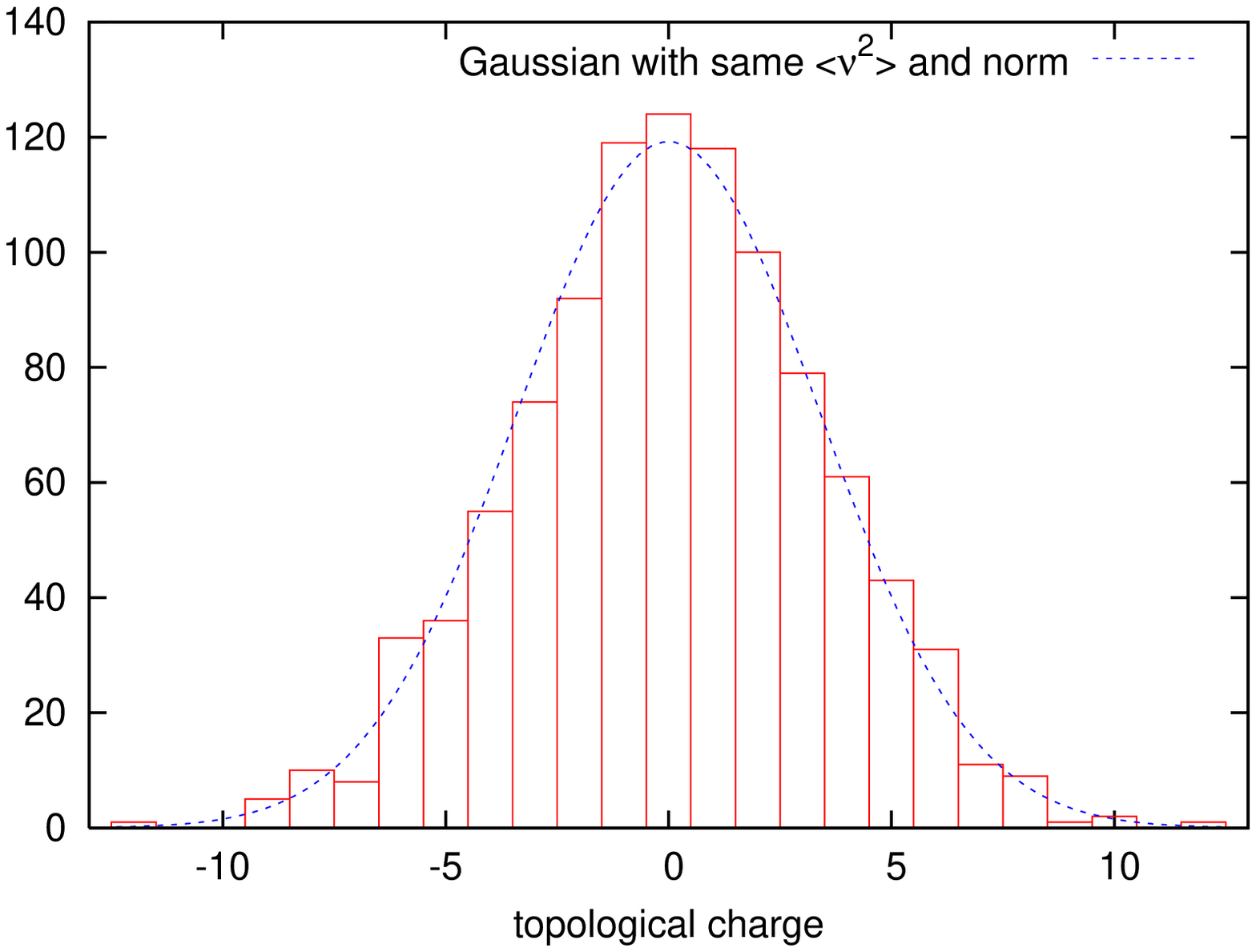}
\caption{{\it The histogram of the overlap-HF indices 
(on the left) and for the standard overlap indices.
In both cases, 1013 indices are included.}}
\label{indhistogram}
\end{figure}

In our study, still at $\beta = 5.85$ on a $12^{3} \times 24$ lattice, 
we compared the charges for the overlap-HF operator described in 
Section 2 and for the standard overlap operator $D_{\rm N}$ at $\mu=1.6$. 
As an example, the histories of about 200 indices for the same configurations
are compared in Fig.\ \ref{tophistory}. Of course, these two types of 
indices are considerably correlated, but only $41 \%$ really coincide. 
We obtained a mean deviation of
\be 
\langle | \nu_{\rm ovHF} - \nu_{\rm N}| \rangle = 0.80(2) \ ,
\ee
and we observed over more than $1000$
configurations a maximal index difference
of $| \nu_{\rm ovHF} - \nu_{\rm N}| =5$. 
Still, the similarity is of
course much closer than the accidental agreement for independent
indices, since they follow essentially the expected Gaussian
distributions, with a width $\approx 3.3$, 
see Fig.\ \ref{indhistogram} and Table \ref{epstab}.
This width fixes the topological susceptibility
\be
\chi_{\rm top} = \frac{1}{V} \langle \nu^{2} \rangle \ ,
\ee
which is of importance to explain the heavy mass of the $\eta'$ meson
\cite{WiVe}. A discussion of that point, as well as a measurement
of $\chi_{\rm top}$ based on $D_{\rm N}$ indices on $L^{4}$ lattices
with a continuum extrapolation, is given in Refs.\ \cite{DDGP}.
 
\begin{figure}[h!]
  \centering
\includegraphics[angle=270,width=.9\linewidth]{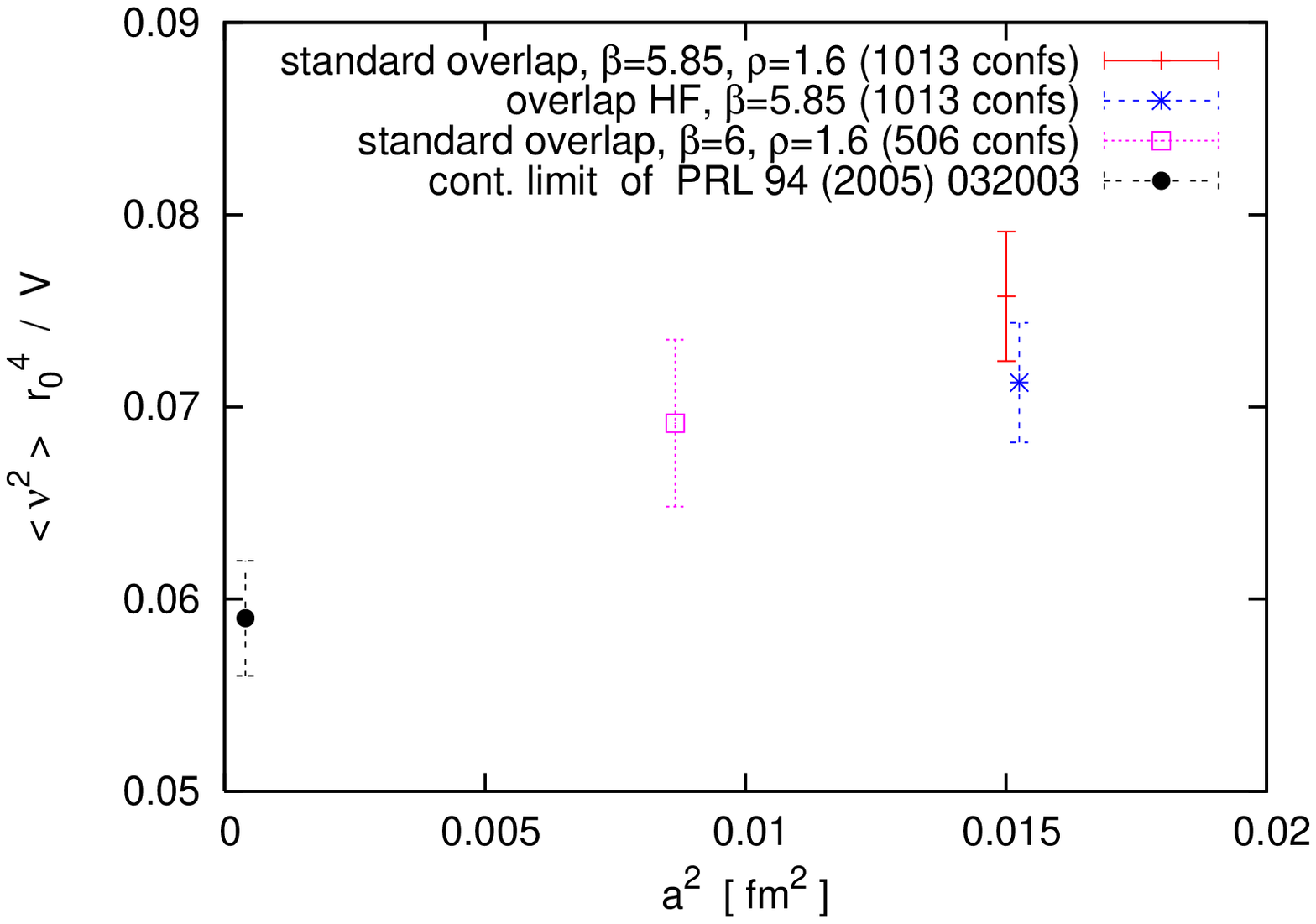}
\caption{{\it The topological susceptibility
measured by indices of $D_{\rm ovHF}$ and of $D_{\rm N}$,
in a volume $V = (1.48~{\rm fm})^{3} \times 
2.96~{\rm fm}$, with two different lattice spacings $a$.
Our data agree well with the
continuum extrapolation reported in Ref.\ \cite{DDGP}.}}
\label{toposus}
\end{figure}

In Fig.\ \ref{toposus} and Table \ref{epstab} 
we present our results with $D_{\rm ovHF}$
and $D_{\rm N}$ on the lattice used so far, plus a result
for $D_{\rm N}$ at $\beta =6$ in the same physical volume
(lattice size $16^{3} \times 32$). We also mark the
continuum extrapolation according to Ref.\ \cite{DDGP}, which is
fully consistent with our results.
This value for $\chi_{\rm top}$ is compatible with the Witten-Veneziano
scenario that much of the $\eta'$ mass is generated by a $U(1)$ anomaly. 

In principle, the charge histogram could give insight
into the possibility of a spontaneously broken parity symmetry in QCD,
which is not fully ruled out \cite{AzGal}. This question was also
studied in Ref.\ \cite{Pisa} with a different definition of topological
charges on the lattice. Here we observe in all cases that the number
of neutral configurations is about half of the corresponding number
with $| \nu |=1$ (see Table \ref{epstab}).
Based on this observation, it cannot be decided if the
charge distribution (at small $| \nu |$) favours a precise Gaussian, 
or a double peak structure (or something else). Hence the fate of 
parity symmetry remains open.

\begin{table}
\begin{center}
\begin{tabular}{|c||c|c|c|}
\hline
Dirac operator & $D_{\rm ovHF}$ & $D_{\rm N}$ & $D_{\rm N}$ \\
\hline
$\beta$ & $5.85$ & $5.85$ & $6$ \\
\hline
$a$ & $\simeq 0.123~{\rm fm}$ & $\simeq 0.123~{\rm fm}$ & $\simeq 0.093~{\rm fm}$ \\
\hline
lattice size & $12^{3}\times 24$ & $12^{3}\times 24$ & $16^{3}\times 32$ \\
\hline
\hline
total \# of confs. & 1013 & 1013 & 506 \\
\hline
$\langle \nu^{2} \rangle$ & $10.81 \pm 0.47$ & $11.49 \pm 0.51$ 
& $10.49 \pm 0.66$ \\
\hline
$\chi_{\rm top} r_{0}^{4}$ & 0.071(3) & 0.076(3) & 0.069(4) \\   
\hline
\hline
\# of confs.\ with $\nu =0$ & 118 & 124 & 59 \\
\hline
\# of confs.\ with $|\nu | =1$ & 238 & 237 & 115 \\
\hline
\# of confs.\ with $|\nu | =2$ & 210 & 192 & 95 \\
\hline
$\Sigma^{1/3}$ \ from sector $\nu =0$ & 298(4) MeV & 301(4) MeV & 279(7) MeV \\
\hline
\hline
$a \langle \lambda_{1} \rangle_{| \nu | =0}$ & 0.0069(4) & 0.0067(4) & 0.0059(4) \\
\hline
$a \langle \lambda_{1} \rangle_{| \nu | =1}$ & 0.0130(4) & 0.0136(4) & 0.0093(3) \\
\hline
$a \langle \lambda_{1} \rangle_{| \nu | =2}$ & 0.0184(5) & 0.0193(5) & 0.0130(5) \\
\hline
$a \langle \lambda_{1} \rangle_{| \nu | =3}$ & 0.0247(6) & 0.0242(8) & 0.0155(6) \\
\hline
$a \langle \lambda_{1} \rangle_{| \nu | =4}$ & 0.0293(9) & ~0.0312(10)& 0.0215(9) \\
\hline
$a \langle \lambda_{1} \rangle_{| \nu | =5}$ & ~0.0338(12)& ~0.0360(13)& ~0.0246(17) \\
\hline
$\Sigma^{1/3}$ from $\langle \lambda_{1} \rangle_{|\nu | = 0 \dots 5}$
& \multicolumn{3}{|c|}{ 290(6) MeV } \\
\hline
\end{tabular}
\end{center}
\caption{{\it Our results for the topological susceptibility
and for the chiral condensate in the $\eps$-regime for a fixed
physical volume $V = (1.48~{\rm fm})^{3} \times 
2.96~{\rm fm}$. We consider two types of overlap operators
($D_{\rm ovHF}$ as described in Section 2, and the standard
overlap operator $D_{\rm N}$ at $\mu =1.6$), and two lattice 
spacings. 
We first give the total statistics
and the resulting topological susceptibility (see Subsection 4.1)
in a dimensionless form, with
$r_{0} = 0.5 ~{\rm fm}$ (according to the Sommer scale \cite{Sommer}). 
\newline
Below we give separately the statistics in the sectors $|\nu | =0,\ 1$
and $2$. We extract the chiral condensate $\Sigma$ from
the density of the lowest Dirac eigenvalue in the
neutral charge sector, which is most reliable
since it involves the smallest values of $z$, see
Subsection 4.2 and Figs.\ \ref{SigmafigovHF}, \ref{SigmafigN}.
\newline
As an alternative we considered the mean values of the lowest
non-zero Dirac eigenvalues $\langle \lambda_{1} \rangle$
in the sectors $|\nu | = 0 \dots 5$. For the value of $\Sigma$ in the
last line {\em all} these results for $\langle \lambda_{1} \rangle_{| \nu|}$
match the RMT predictions, see Fig.\ \ref{l1mean}.}}
\label{epstab}
\end{table}

\subsection{Determination of the chiral condensate $\Sigma$}

Chiral RMT conjectures predictions for the low
lying eigenvalues, ordered as $\lambda_{n}$, $n = 1,2, \dots$
(excluding possible zero eigenvalues)
of the Dirac operator in the $\epsilon$-regime
(for a review, see Ref.\ \cite{VerWet}).
More precisely, the conjectured densities are functions of
the dimensionless variables $\Sigma V \lambda_{n}$,
where $\Sigma$ is the chiral condensate in the effective 
Lagrangian (\ref{Leff}). Here we focus on
the variable $z := \Sigma V \lambda_{1,P}$,
where $\lambda_{1,P}$ emerges from the leading non-zero eigenvalue
$\lambda_{1}$ if the spectral circle of the overlap operator
is mapped stereographically on the imaginary axis,
$\lambda_{1,P} = | \lambda_{1} / (1 - a \lambda_{1} / 2 \mu )|$.
\footnote{Alternatively, one could simply consider $|\lambda_{1}|$,
which are eigenvalues of $\gamma_{5}D$, but for the small eigenvalues
that we deal with the difference is not of importance.}

\begin{figure}[hbt]
  \centering
\includegraphics[angle=270,width=.8\linewidth]{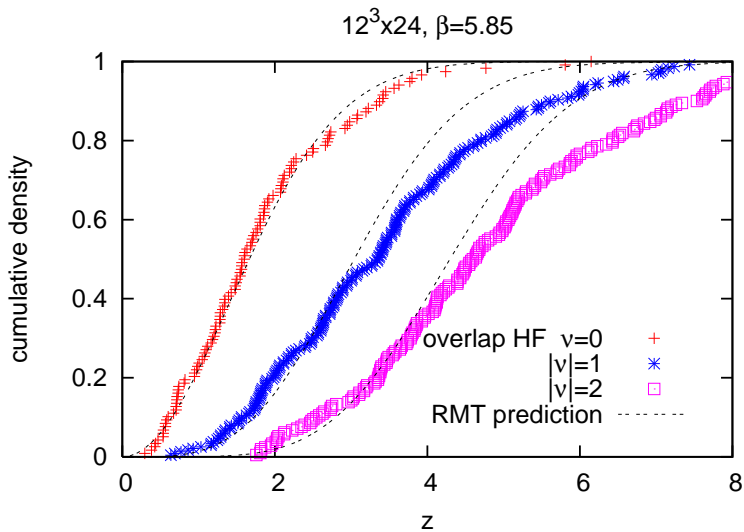}
\caption{{\it The cumulative density of the lowest Dirac eigenvalue 
$\lambda_{1,P}$ of the overlap-HF operator, 
in the topological sectors $|\nu | =0$, $1$ and $2$. 
We compare the chiral RMT predictions to our data
for $z= \Sigma V \lambda_{1,P}$
with $\Sigma^{1/3} = 298 ~ {\rm MeV}$ --- the optimal value in the
neutral sector ($\nu = 0$). This value works well up to $z \approx 3$
for all topological sectors.}}
\label{SigmafigovHF}
\end{figure}

\begin{figure}[h!]
  \centering
\includegraphics[angle=270,width=.8\linewidth]{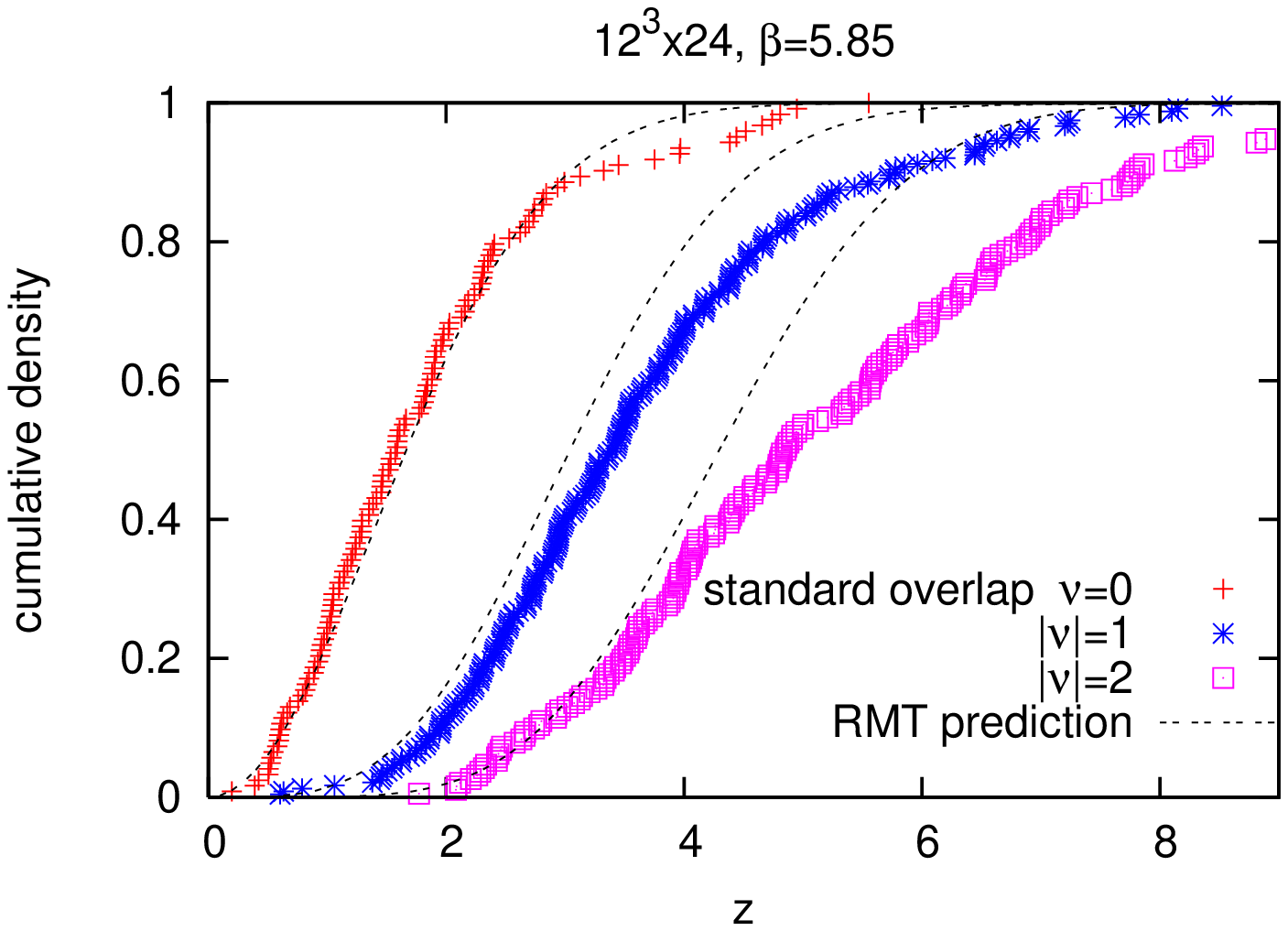}
\includegraphics[angle=270,width=.8\linewidth]{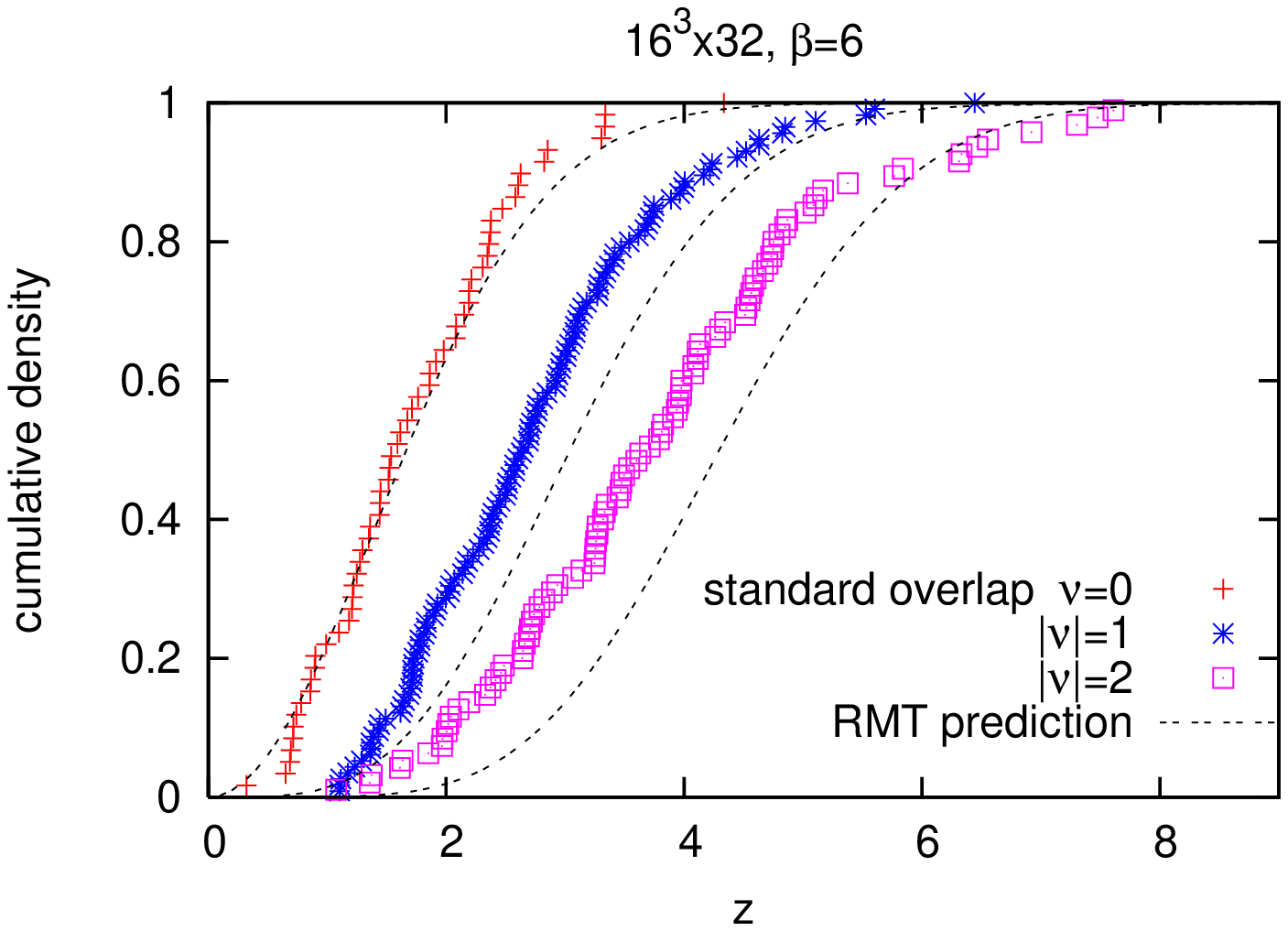}
\caption{{\it The cumulative density the lowest Dirac eigenvalue 
$\lambda_{1,P}$ of the standard overlap operator $D_{\rm N}$ at 
$\beta =5.85$ (on top) and $\beta =6$ (below), 
in the topological sectors $|\nu | =0$, $1$ and $2$. 
We compare the chiral RMT predictions to our data
for $z= \Sigma V \lambda_{1,P}$
with $\Sigma^{1/3} = 301 ~ {\rm MeV}$ (on top),
and with $\Sigma^{1/3} = 279 ~ {\rm MeV}$ (below)
--- the optimal values in the neutral sector ($\nu = 0$).
Considering also $| \nu | =1,2$ would decrease (increase) $\Sigma$ at
$\beta =5.85$ ($\beta =6$) which is the trend towards the result
with the method of Fig.\ \ref{l1mean}.}}
\label{SigmafigN}
\end{figure}

These RMT predictions depend on $| \nu |$, the absolute
value of the topological charge. Here we make use
of the explicit formulae \cite{lowEV} for the density of the first
non-zero (re-scaled) eigenvalues in the sectors
$|\nu |$, which we denote by $\rho_{1}^{(| \nu |)}(z)$.
For the lowest eigenvalues, the particular density 
$\rho_{1}^{(0)}$ was first confirmed by staggered fermion simulations
(results are summarised in Ref.\ \cite{VerWet}), 
but the charged sectors yielded the very same density,
in contradiction to RMT. The distinction between the topological
sectors was first observed to hold for $D_{\rm N}$
to a good precision \cite{RMT}, if the linear box 
size exceeds a lower limit of about $L \gsim 1.1~{\rm fm}$ 
(of course, the exact limit depends  on the criterion).
\footnote{Meanwhile, a topological splitting was also observed
to set in for staggered fermion if the link variables are strongly smeared
\cite{staggtop}.}
Once the predicted density $\rho_{1}^{(| \nu |)}$ is well reproduced,
we can read off the value of $\Sigma$, which is the only free
fitting parameter for all topological sectors.
It is most instructive to plot the cumulative
densities \cite{NR}, which we show in Figs.\ \ref{SigmafigovHF}
and \ref{SigmafigN}. We compare here the predictions to
the eigenvalues $\lambda_{1,P}$, which we measured in various
topological sectors.
The statistics involved in each case is included in Table \ref{epstab}.
This Table also displays the $\Sigma$ values obtained in the
sector $\nu = 0$, which we consider most reliable, since it deals
with the lowest eigenvalues resp.\ energies.
As a theoretical bound, one often refers to the Thouless energy
$F_{\pi}^{2} /( \Sigma \sqrt{V})$, below which these predictions should 
hold. In our case, it translates into $z_{\rm Thouless} \lsim 1$,
but the eigenvalue distributions follow the chiral RMT behaviour
up to larger $z$ values.
Clearly, in this range it is the neutral
sector ($\nu =0$) which contributes in a dominant way, but
Fig.\ \ref{SigmafigN} shows that in the case of $D_{\rm N}$
(for both values of $\beta$)
the charged sectors $| \nu | =1$ and $2$ alone would favour a different 
$\Sigma$ value. This ambiguity also occurs for smeared staggered
fermions \cite{staggtop}. In the case of the $D_{\rm ovHF}$,
however, a unique $\Sigma$ works well for all the three sectors
$| \nu | = 0,1,2$, up to about $z \approx 3$,
as we see from Fig.\ \ref{SigmafigovHF}.

\begin{figure}[h!]
  \centering
\includegraphics[angle=270,width=.9\linewidth]{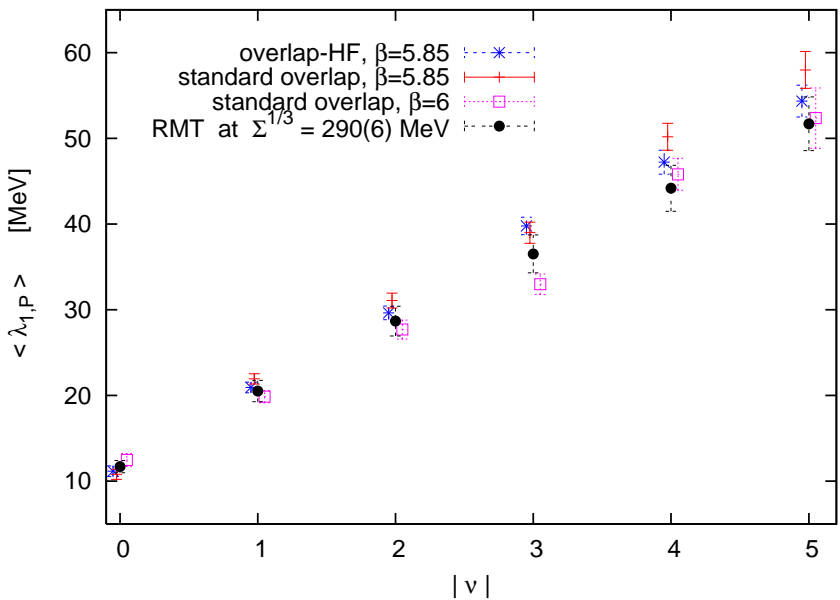}
\caption{{\it The mean values of the first non-zero Dirac eigenvalue
in the charge sectors $|\nu | =0 \dots 5$, in physical unit.
All the measured results agree with chiral RMT if we choose
$\Sigma^{1/3} = 290(6) ~ {\rm MeV}$.}}
\label{l1mean}
\end{figure}

As an alternative approach to test the agreement of our data
with the chiral RMT, and to extract a value for $\Sigma$, we now focus
on the mean values of the leading non-zero Dirac eigenvalues $\lambda_{1}$
in all the charge sectors up to $| \nu |=5$. In physical units,
the results $\langle \lambda_{1,P} \rangle$ agree remarkably well
for the different overlap operators and lattice spacings
--- see Fig.\ \ref{l1mean} ---
although this consideration extends beyond very low energy. Each single
result for $\langle \lambda_{1,P} \rangle_{| \nu |}$ 
can then be matched to the RMT value for a specific choice of
$\Sigma$. Amazingly, all these 18 results are in agreement
with the RMT if we fix
\begin{equation}
\Sigma = (290(6) ~ {\rm MeV})^{3} \ ,
\end{equation}
as Fig.\ \ref{l1mean} also shows.
This value is between the results obtained from the eigenvalue densities
at $\nu =0$ alone, and we recognise from Figs.\ \ref{SigmafigovHF} and
\ref{SigmafigN} that this is the trend if we take the charged
sectors into account.

A renormalisation procedure for $\Sigma$ obtained in this
way is discussed in Ref.\ \cite{WenWit}. However, we will only use 
this quenched lattice result as a fitting
input in Section 4.3, so here we stay with the bare condensate
$\Sigma$ for our fixed lattice parameters.

\subsection{Evaluation of $F_{\pi}$ based on the axial-vector current
correlator}

As we mentioned before, QCD simulations with chiral quarks can only 
be performed in the quenched approximation for the time being.
In order to relate simulation results to the effective low energy theory,
we therefore refer to {\em quenched $\chi$PT}. In that framework,
mesonic correlation functions were evaluated to the first order in
Refs.\ \cite{corre1,corre2}. It turned out that the vector current
correlation function vanishes; this property actually extends to all orders 
\cite{corre3}. The scalar and pseudoscalar correlators involve
already in the first order additional, quenching specific LECs,
which obstruct the access to the physical LECs in the
Lagrangian (\ref{Leff}). Therefore we first focus on the axial-vector
correlator, which only depends on $\Sigma $ and $F_{\pi}$ in the
first order (as in the dynamical case). In particular we are going
to compare our data to the quenched $\chi$PT prediction
in a volume $V= L^{3} \times T$ \cite{corre2},
\bea   \label{AA}
Z_{A}^{2} \cdot \la A_{4}(t) \, A_{4}(0) \ra_{\nu} &=& 
2  \left(\frac{F_{\pi}^{2}}{T} 
 + 2 m_{q} \, \Sigma_{\vert \nu \vert}(z_{q}) \, 
T  \, h_{1}(\tau )\right) \ , \\
h_{1}(\tau ) &=& \frac{1}{2} \Big( \tau^{2} - \tau + \frac{1}{6} \Big) 
\ , \qquad \tau = \frac{t}{T} \ , \nn \\
\Sigma_{\nu}(z_{q}) &=& \Sigma \left( z_{q} \Big[ I_{\nu}(z_{q}) K_{\nu}(z_{q}) + 
I_{\nu +1}(z_{q}) K_{\nu -1}(z_{q}) \Big] + \frac{\nu}{z_{q}} \right) \nn \ ,
\eea
where
\be
A_{4}(t) = a^{3} \sum_{\vec x} \bar \psi (t ,\vec x) \gamma_{5}
\gamma_{4} \psi (t , \vec x ) \qquad \quad (t > 0)
\ee
is the bare axial-vector current at 3-momentum $\vec p = \vec 0$.
This formula applies to the topological sectors of charge
$\pm \nu$. 
$I_{\nu}$ and $K_{\nu}$ are modified Bessel functions, and
$z_{q} := \Sigma V m_{q}$ (in analogy to the variable $z$
in Subsection 4.2).

It is remarkable that this prediction in the 
$\epsilon$-regime has the shape of a {\em parabola} with a minimum
at $t = T/2$. This is in qualitative contrast to the ${\tt cosh}$
behaviour, which is standard in large volumes. $\Sigma$
affects both, the curvature and the minimum of this parabola,
whereas $F_{\pi}$ only occurs in the additive constant --- that
feature is helpful for its evaluation.

A first comparison of this curve to lattice data was presented
in Ref.\ \cite{AApap}, using $D_{\rm N}$ at $\beta =6$, 
$am_{q} = 0.01$ on lattice volumes $10^{3} \times 24$
and $12^{4}$. The first among these volumes --- with a linear
size of $L \simeq 0.93~{\rm fm}$ --- turned out to be too small:
the data for $\la A_{4}(t) \, A_{4}(0) \ra_{1,2}$ were practically
flat in $t$ and incompatible with the parabola of eq.\ (\ref{AA})
for any positive $\Sigma$.
This observation was consistent with the lower bound for $L$ that we
also found for the agreement of the microscopic spectrum with chiral RMT. 

Another observation in that study was that the corresponding history
in $\nu =0$ is plagued by strong spikes, giving rise to large
statistical errors. A huge statistics
($O(10^{4})$ topologically neutral configurations) 
would be required for conclusive results
(see also Ref.\ \cite{nuzero}). These spikes occur for the configurations
with a tiny (non-zero) Dirac eigenvalue $\lambda_{1,P}$, 
and it agrees again with chiral
RMT that such configurations are most frequent in the topologically
neutral sector.
As a remedy to this problem, a method called ``low mode averaging''
was designed \cite{LMA}. 

However, without applying that method we obtained a decent agreement
with the prediction (\ref{AA}) in our second volume mentioned above
($V \simeq (1.12 ~{\rm fm})^{4}$) in the sector
$| \nu | = 1$, and still a reasonable shape --- although somewhat flat ---
at $| \nu | = 2$ \cite{AApap}. In view of the leading LECs, it seems 
unfortunately impossible to extract a value of $\Sigma$ from such data, since 
the theoretical curvature depends on it only in an extremely weak way
(for instance, even an extreme change from $\Sigma =0$ to $\Sigma
= (250~ {\rm MeV})^{3}$ had such a small effect that it is practically
hopeless to resolve it from lattice data).
\footnote{Only in the sector $\nu =0$ the sensitivity to $\Sigma$ is
significant, but there we run into the statistical problem
mentioned before.}

On the other hand, $F_{\pi}$ can be extracted quite well from the 
vicinity of the minimum at $t = T/2$, but the value found in Ref.\
\cite{AApap} was too high.
\footnote{In Ref.\ \cite{AApap} we gave a value around $130~{\rm MeV}$,
but the analysis did not handle the renormalisation constant $Z_{A}$
very carefully --- being more precise in this aspect reduces the
value to about $120~{\rm MeV}$.}

Next a study of that kind appeared in Ref.\ \cite{Japs},
which also used $D_{\rm N}$, at $\beta = 5.85$ and $\mu =1.6$, now
on a $10^{3} \times 20$ lattice. These authors analysed
the sectors $| \nu |=0$
and 1 (without ``low mode averaging'') and arrived at
$F_{\pi } = (98.3 \pm 8.3) ~ {\rm MeV}$. As a reason for
the limitation to $| \nu | \leq 1$, Ref.\ \cite{Japs} refers to the
condition $| \nu | \ll \langle \nu^{2} \rangle$. As we mentioned in
Subsection 4.2, one expects $\langle \nu^{2} \rangle \propto V$
(up to lattice artifacts), hence this limitation was imposed by the
volume.

\begin{figure}[h!]
  \centering
\includegraphics[angle=270,width=.67\linewidth]{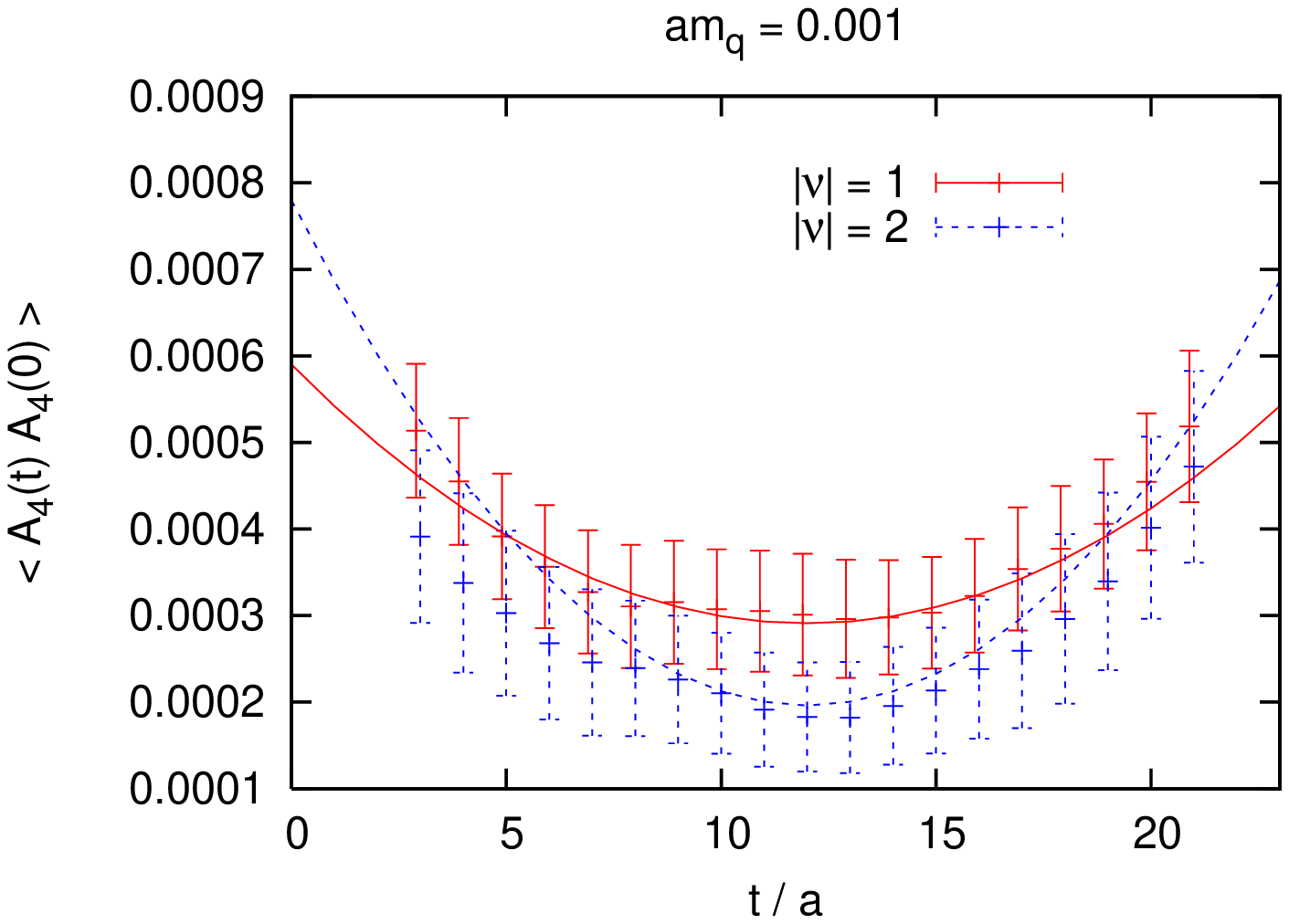} 
\includegraphics[angle=270,width=.67\linewidth]{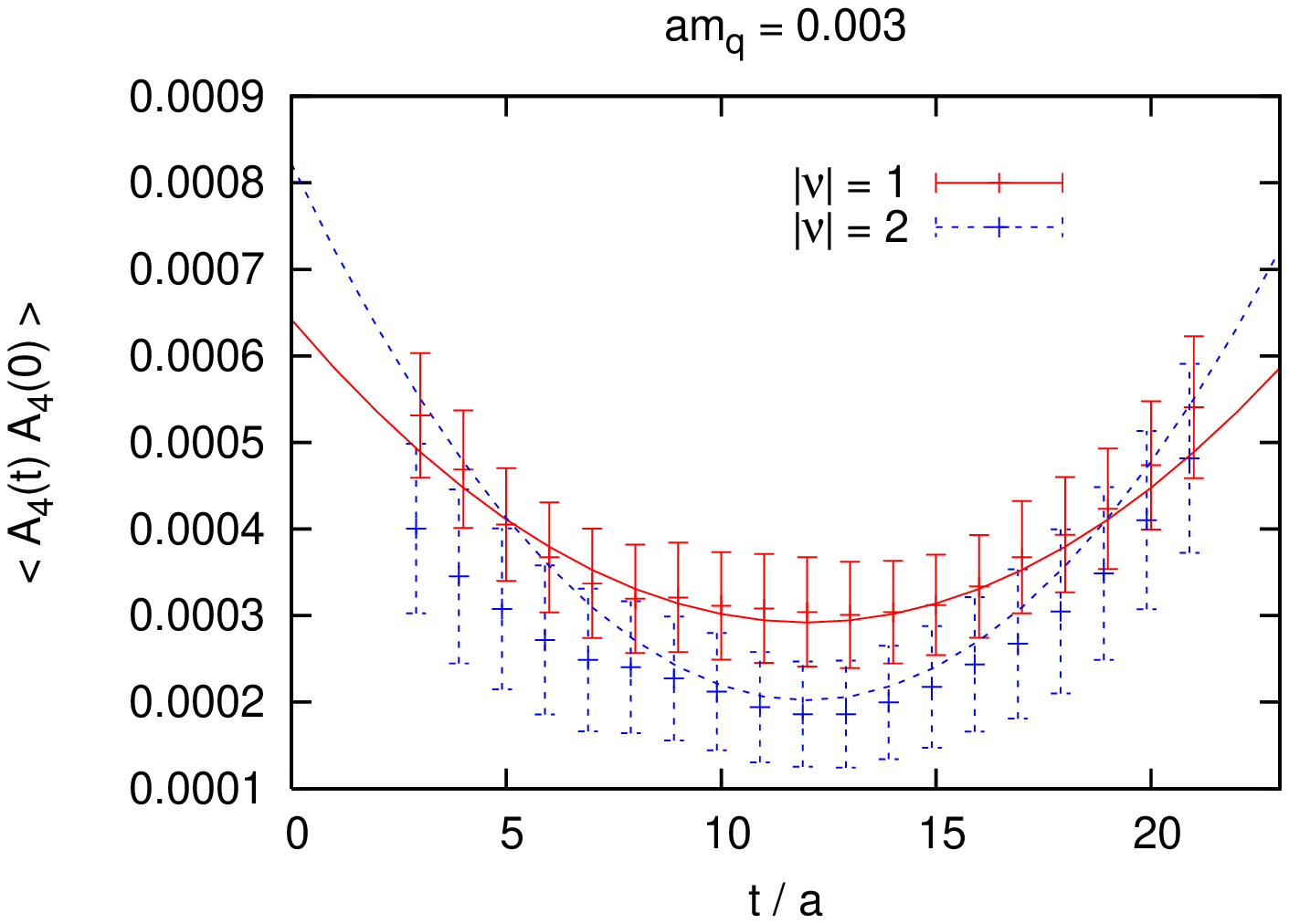}
\includegraphics[angle=270,width=.67\linewidth]{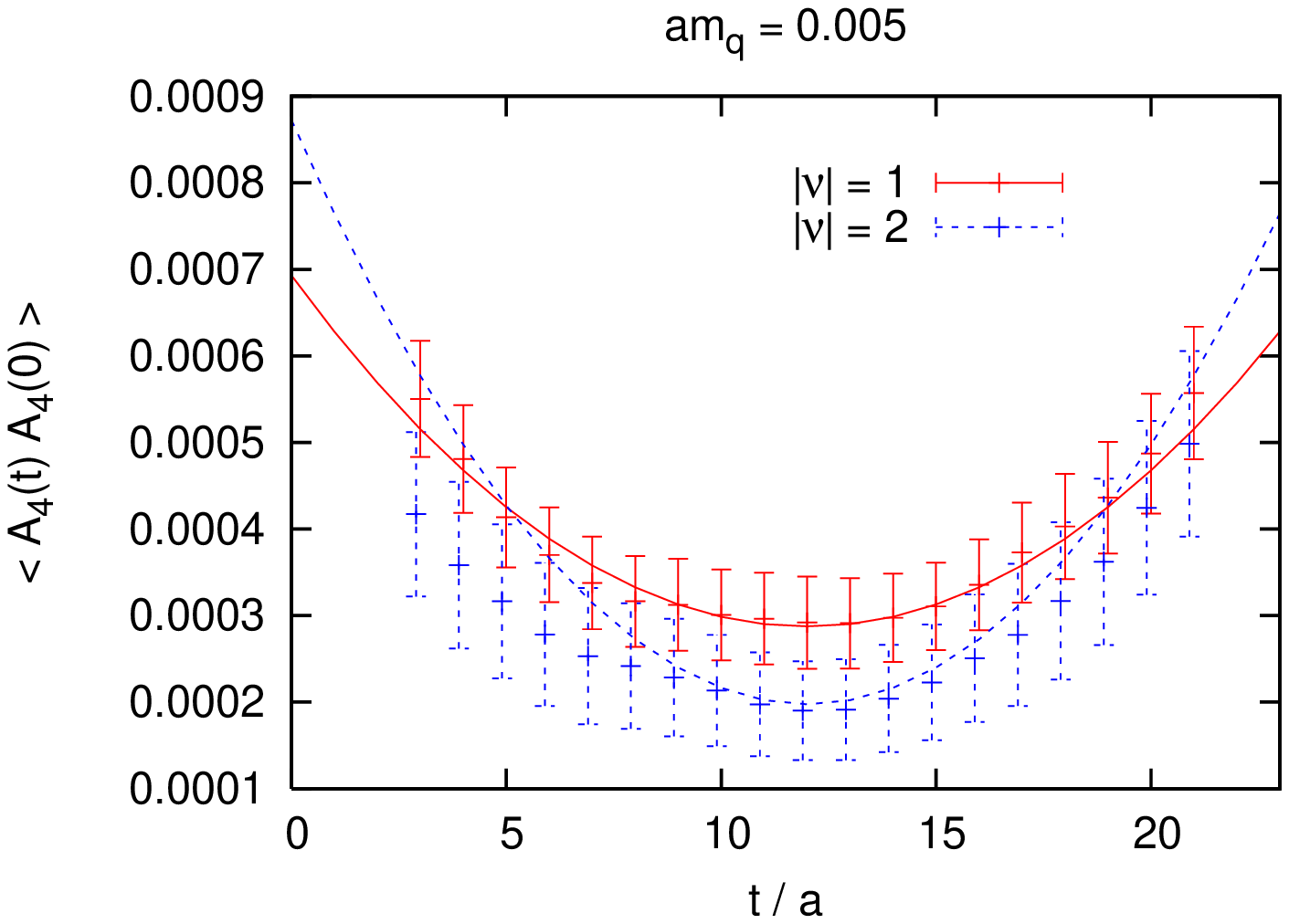}
\caption{{\it Lattice data with $D_{\rm ovHF}$ 
vs.\ predictions by quenched $\chi$PT
for the axial-current correlation functions in the $\epsilon$-regime,
measured separately in the topological sectors $| \nu | = 1$ and $2$.
The global fit at each mass corresponds to 
the values of $F_{\pi}$ given in Table \ref{epstabFpiAA}.}}
\label{AAfig}
\end{figure}

\begin{figure}[ht!]
  \centering
\includegraphics[angle=270,width=.67\linewidth]{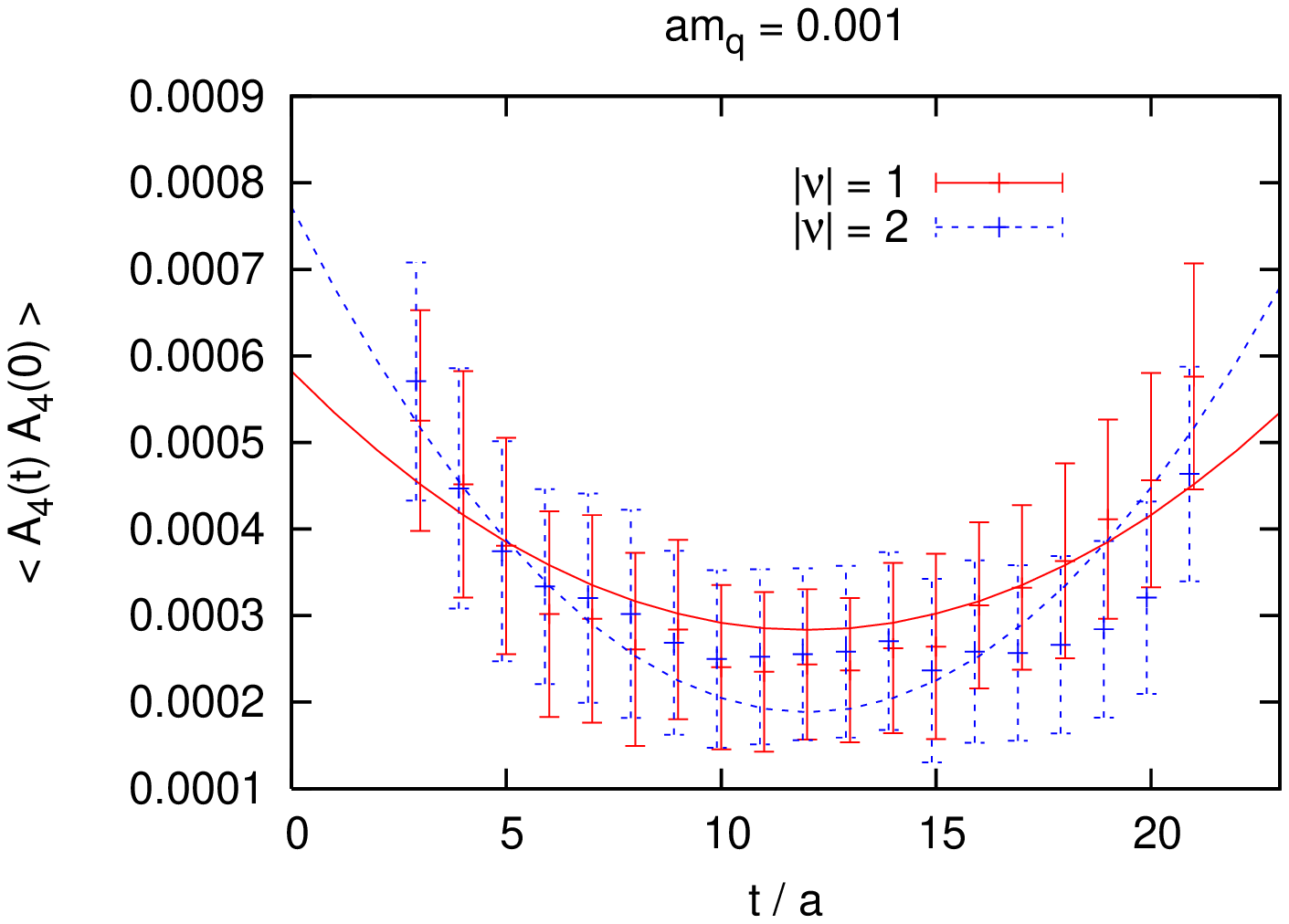}
\includegraphics[angle=270,width=.67\linewidth]{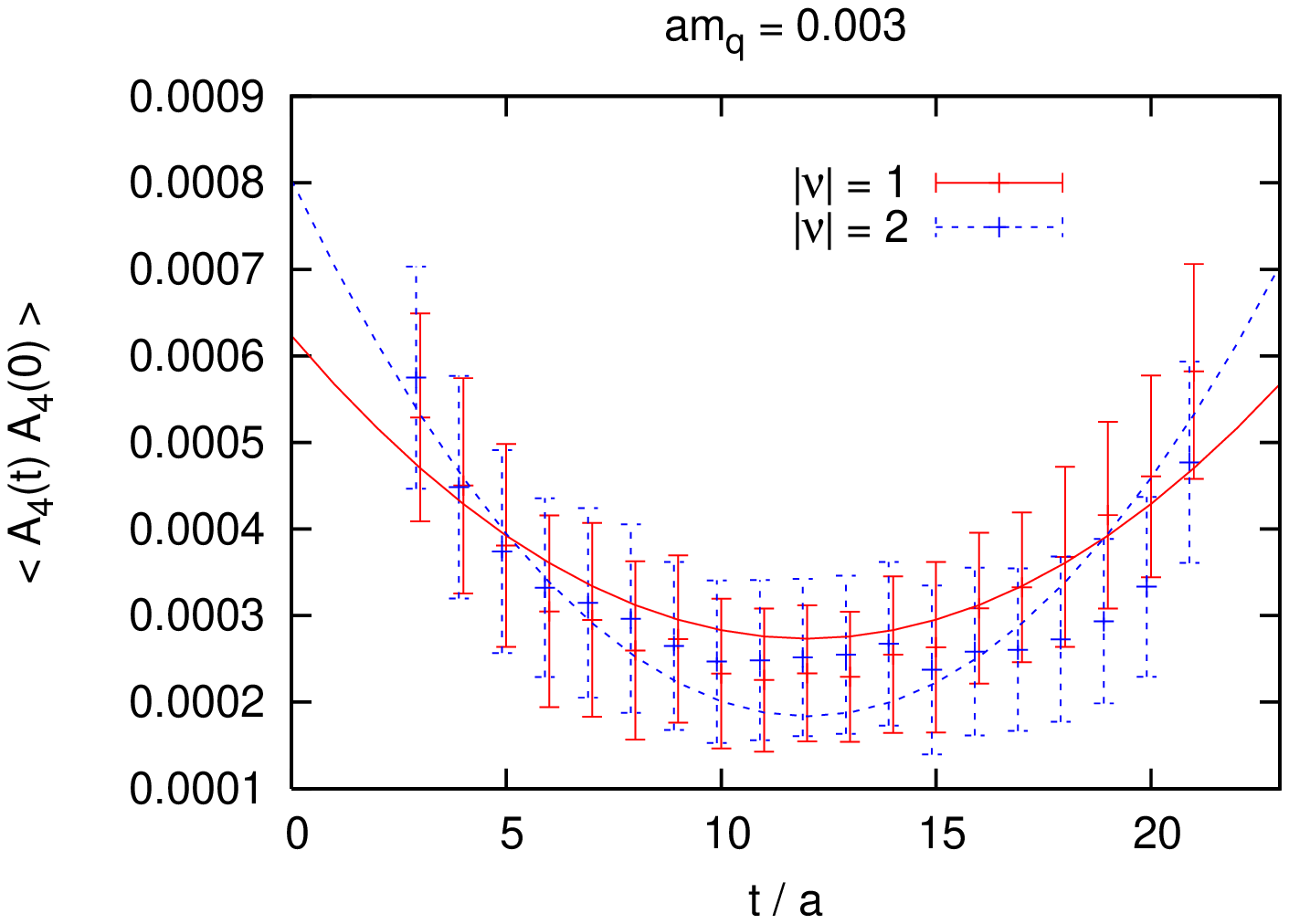} 
\includegraphics[angle=270,width=.67\linewidth]{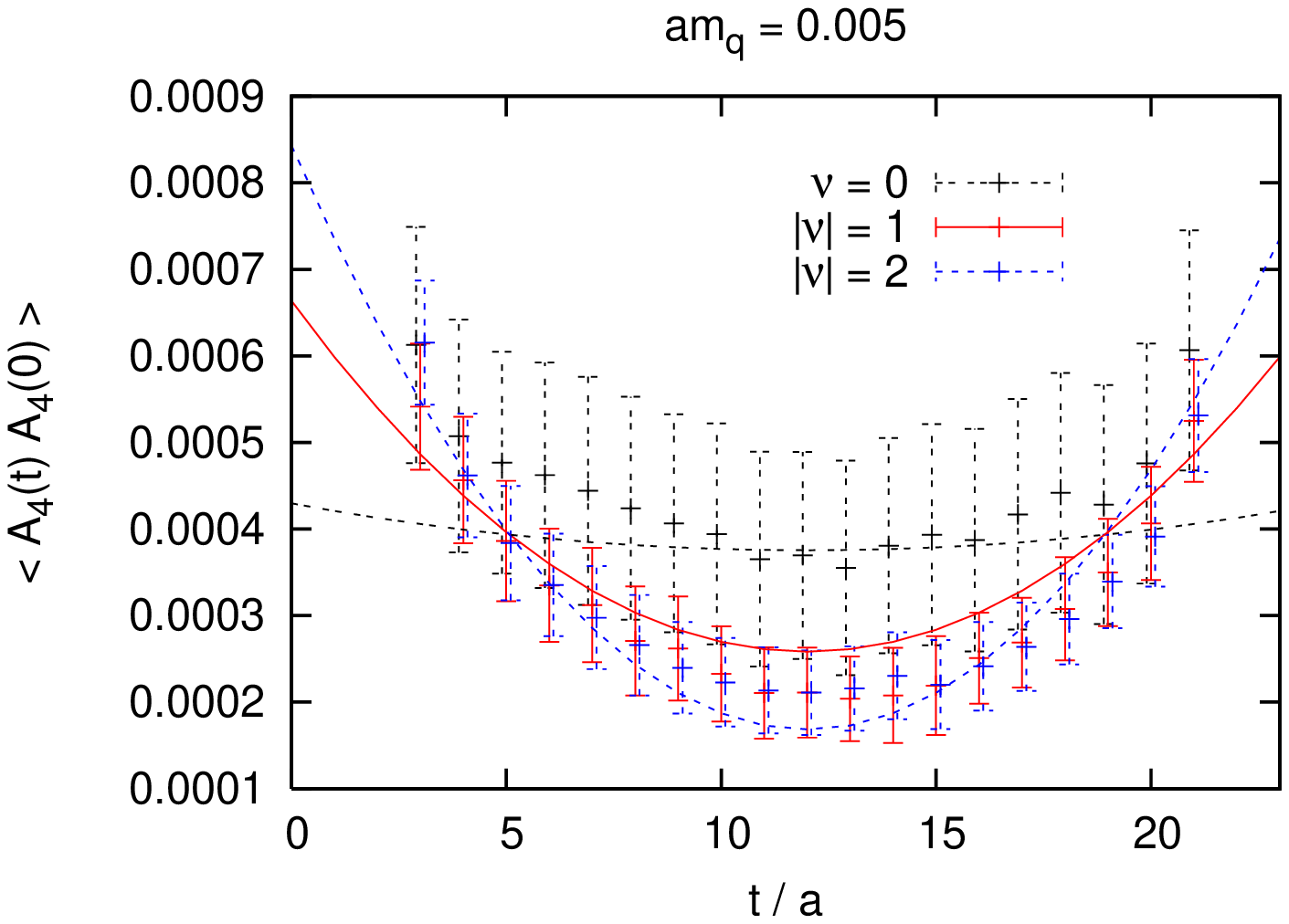}
\caption{{\it Lattice data with $D_{\rm N}$ 
vs.\ predictions by quenched $\chi$PT
for the axial-current correlation functions in the $\epsilon$-regime,
measured separately in the topological sectors $| \nu | = 0$, $1$ 
and $2$. The global fit at each mass corresponds to the values 
of $F_{\pi}$ given in Table \ref{epstabFpiAA}.}}
\label{AA_Nfig}
\end{figure}

Here we present again results at $\beta = 5.85$ on a $12^{3} \times 24$
lattice, where also the latter condition 
admits $| \nu | = 2$, c.f.\ Table \ref{epstab}.
We evaluated for both, $D_{\rm ovHF}$ and $D_{\rm N}$,
the axial-vector correlators at $am_{q}=0.001$, $0.003$ and
$0.005$, which turns out to be safely in the $\epsilon$-regime.
Our propagator statistics at these quark masses is given in
Table \ref{epstabFpiAA}. We then fitted the data to eq.\ (\ref{AA})
by using the chirally extrapolated factors $Z_{A}$ ($1.17$ for
$D_{\rm ovHF}$ and $1.45$ for $D_{\rm N}$), along with the
$\Sigma$ values that we obtained from the microscopic Dirac spectra
(see Subsection 4.2 and Table \ref{epstab}).
For each of the overlap operators we performed at each of the quark
masses a global fit over the topological
sectors that we considered, which is shown in Figs.\ \ref{AAfig}
and \ref{AA_Nfig}. 
In particular our results for $D_{\rm ovHF}$ reveal for the first time
a quite clear distinction between the sectors $| \nu | = 1$ and
$| \nu | = 2$ --- this property could not be observed for $D_{\rm N}$
up to now. For $D_{\rm N}$ at $am_{q} =0.005$ we also include
the neutral sector; as expected it has by far larger error bars than the 
charged sectors, but it is helpful nevertheless to reduce the 
error on $F_{\pi}$ in the global fit.
\begin{table}
\begin{center}
\begin{tabular}{|c||c|c|}
\hline
Dirac operator & $D_{\rm ovHF}$ & $D_{\rm N}$ \\
\hline
\hline
$a m_{q} = 0.001$ && \\
\hline
\# of propagators at $|\nu | =1$ & 50 & 50 \\
\hline
\# of propagators at $|\nu | =2$ & 50 & 50 \\
\hline
$F_{\pi}$ & $(110 \pm 8) ~ {\rm MeV}$ & $(109 \pm 11) ~ {\rm MeV}$ \\
\hline
\hline
$a m_{q} = 0.003$ && \\
\hline
\# of propagators at $|\nu | =1$ & 50 & 50 \\
\hline
\# of propagators at $|\nu | =2$ & 50 & 50 \\
\hline
$F_{\pi}$ & $(113 \pm 7) ~ {\rm MeV}$ & $(110 \pm 11) ~ {\rm MeV}$ \\
\hline
\hline
$a m_{q} = 0.005$ && \\
\hline
\# of propagators at $\nu = 0$ & -- & 100 \\
\hline
\# of propagators at $|\nu | =1$ & 50 & 100 \\
\hline
\# of propagators at $|\nu | =2$ & 50 & 100 \\
\hline
$F_{\pi}$ & $(115 \pm 6) ~ {\rm MeV}$ & $(111 \pm 4) ~ {\rm MeV}$ \\
\hline
\end{tabular}
\end{center}
\caption{{\it Our results in the $\eps$-regime for the pion
decay constant $F_{\pi}$, based on the axial-current correlation function.
The results are obtained at $\beta = 5.85$ on a $12^{3} \times 24$ lattice.
We give our statistics of the propagators in the different topological
sectors at various bare quark masses in the $\epsilon$-regime. 
(Of course, different configurations where used for $D_{\rm ovHF}$
and for $D_{\rm N}$.)
The results for $F_{\pi}$ were determined from fits to the quenched
$\chi$PT formula (\ref{AA}) in the range \ $t / a \in [11,13]$,
see Figs.\ \ref{AAfig} and \ref{AA_Nfig}.}}
\label{epstabFpiAA}
\end{table}

The emerging values for the pion decay constant are consistent,
\bea
F_{\pi} &=& (110 \pm 6) ~ {\rm MeV} \qquad ({\rm using} ~ 
D_{\rm ovHF}) \ ,  \nonumber \\
F_{\pi} &=& (109 \pm 4) ~ {\rm MeV} \qquad ({\rm using} ~ D_{\rm N}) \ .
\eea

\begin{figure}[ht!]
  \centering
\includegraphics[angle=270,width=.67\linewidth]{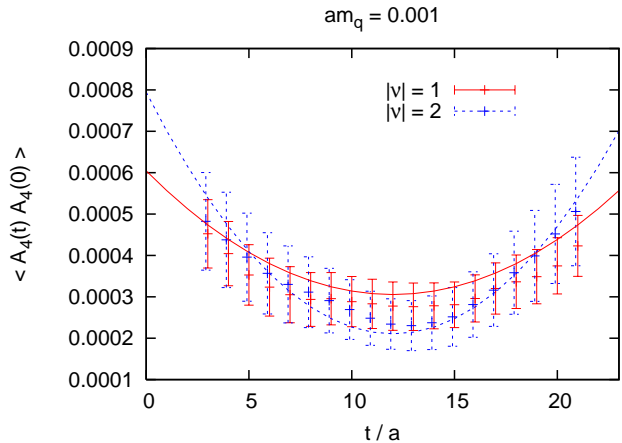}
\caption{{\it Lattice data with $D_{\rm ovHF}$ 
vs.\ predictions by quenched $\chi$PT
for the axial-current correlation functions, as in Fig.\ \ref{AAfig},
but here the data are re-weighted with the first 
non-zero Dirac eigenvalue, i.e.\ with the factor (\ref{reweight}).
This re-weighting --- which is most powerful at minimal $m_{q}$ ---
improves the agreement with the theoretical prediction at 
relatively large $|t - T/2|$.
In this case, the fit leads to $F_{\pi} = (112 \pm 7) ~ {\rm MeV}$.}}
\label{AA_rewe}
\end{figure}

As an experiment, we also considered a simple re-weighting of the
axial-current correlators involved, by means of the factor
\be  \label{reweight}
m_{q}^{|\nu |} \, \Big[ m_{q}^{2} + 
\Big(1 - \frac{a m_{q}}{2 \mu}\Big)^{2} \lambda_{1,P}^{2} \Big] \ ,
\ee
which is (to a very good approximation) part of the fermion determinant.
Since we take the statistics inside fixed sectors,
the factor $m_{q}^{|\nu |}$ does not matter here, but the second factor
attaches weights to the contributions, which differ in particular
for very small $m_{q}$. As an example, we show in Fig.\ \ref{AA_rewe} 
the result obtained in this way for the overlap-HF at $am_{q}=0.001$.
Of course, this is a modest step towards a $1$ flavour re-weighting, which
works well in some cases if a few hundred low lying eigenvalues are involved
\cite{Japs,rewei}. Still, in the present case we observe for the
overlap-HF data an improved agreement with the predicted curves
for $| \nu | =1$ and $2$ at $t$ values relatively far from $T/2$.


\subsection{Evaluation of $F_{\pi}$ based on the zero-modes}

At last we still consider an alternative method to evaluate
$F_{\pi}$ in the $\epsilon$-regime. 
This method was introduced in Ref.\ \cite{zeromodes}, 
and it involves solely the zero-mode contributions to the
pseudoscalar correlation function. Hence we now work directly
in the chiral limit. Let us briefly summarise
the main idea of this approach.


Ref.\ \cite{zeromodes} computed
the chiral Lagrangian to the next-to-next-to-lead\-ing order
in quenched $\chi$PT, ${\cal L}_{q\chi PT}^{(2)}$.
It can be written in a form that involves an auxiliary scalar field
$\Phi_{0}$, which is coupled to the quasi Nambu-Goldstone field $U$
by a new LEC denoted as $K$.
The auxiliary field also contributes
\be
{\cal L}^{(2)} [ \Phi_{0} ] =
\frac{\alpha_{0}}{2 N_{c}} \partial_{\mu} \Phi_{0} \partial_{\mu} \Phi_{0} 
+ \frac{m_{0}^{2}}{2 N_{c}} \Phi_{0}^{2}
\ee
to ${\cal L}_{q\chi PT}^{(2)}$, which brings in $\alpha_{0}$ and $m_{0}$
as another two quenching specific LECs, in addition to $K$.
The inclusion of $\Phi_{0}$ supplements the quenching effects;
in the dynamical case it decouples form the Nambu-Goldstone field.

It is somewhat ambiguous how to count these additional terms in the
quenched $\epsilon$-expansion. Ref.\ \cite{zeromodes} assumes
the action terms with the coefficients $\alpha_{0}$ and $ \ K \sqrt{N_{c}}$ 
to be of $O(1)$, whereas the one with $m_{0}$ is in $O(\epsilon )$.
In particular the last assumption is a bit unusual; 
for instance, it disagrees with the framework referred to in Subsection
4.3. However, it is an acceptable possibility, which simplifies
this approach since it removes
the auxiliary mass term from the dominant order. If one further
defines the dimensionless parameter
\be
\alpha = \alpha_{0} - \frac{4 N_{c}^{2} K F_{\pi} }{\Sigma} \ ,
\ee
then only the LECs $F_{\pi}$ and $\alpha$ occur in this order.

For $N_{f}$ valence quark flavours, we now consider the correlation
function of the pseudoscalar density $P(x)$, which can be decomposed into
a connected plus a disconnected part,
\bea
V^{2} \langle P(x) P(y) \rangle &=& N_{f} P_{1}(x,y) - N_{f}^{2} P_{2}(x,y) \\
P_{1}(x,y) &=& {\rm Tr} [ i \gamma_{5} (D + m_{q})^{-1} (x,y) \cdot
i \gamma_{5} (D + m_{q})^{-1} (y,x) ] \nn \\ 
P_{2}(x,y) &=& {\rm Tr} [ i \gamma_{5} (D + m_{q})^{-1} (x,x) \cdot
i \gamma_{5} (D + m_{q})^{-1} (y,y) ] \ . \quad \nn
\eea
Then one performs a spectral decomposition of the propagators and obtains
the residuum in terms of the zero-modes,
\bea
^{~~ \lim}_{m_{q} \to 0} \, (m_{q}V)^{2} \langle P(x) P(0) \rangle_{\nu} &=&
N_{f} C^{(1)}_{| \nu |}(x) + N_{f}^{2} C^{(2)}_{| \nu |}(x) \nn \\
{\rm connected:~~~}
C^{(1)}_{| \nu |}(x) &=& - \langle v_{j}^{\dagger} (x) v_{k}(x) \cdot
v_{k}^{\dagger} (0) v_{j}(0) \rangle_{| \nu |} \nn \\
{\rm disconnected:~~~}
C^{(2)}_{| \nu |}(x) &=& \langle v_{j}^{\dagger} (x) v_{j}(x) \cdot
v_{k}^{\dagger} (0) v_{k}(0) \rangle_{| \nu |} \ .
\eea
The vectors $v_{j}$ denote the (exact) zero-modes of the
Ginsparg-Wilson operator at mass zero, $D_{\rm GW} \, v_{j}=0$.
In the terms for $C^{(i)}_{| \nu |}$ the zero-modes are summed over.

Next we consider the spatial integral
$\int d^{3}x \, P(x) P(0)$. Now the above procedure for the correlation function
leads to functions $C^{(i)}_{| \nu |} (t)$, $i=1,2$, which are given explicitly
in Ref.\ \cite{zeromodes}. In principle, these functions could be measured
and fitted to the predictions in order to determine $F_{\pi}$ and $\alpha$.
In practice, however, it is much better to consider instead just 
the leading term in the expansion at $t = T/2$,
\be  \label{Taylor}
\frac{V}{L^{2}} \frac{d}{dt} C^{(i)}_{| \nu |} (t) |_{t = T/2} =
D^{(i)}_{| \nu |} s + O(s^{3}) \ , \quad s = t - \frac{T}{2} \ , 
\quad i = 1,2 \ .
\ee
The slopes $D^{(i)}_{| \nu |}$ tend to
be stable over a variety of fitting ranges 
$s \in [ - s_{\rm max}, s_{\rm max} ]$,
$s_{\rm max} = a,\, 2a ,\, 3a \dots $. 
To be explicit, the slope functions \cite{zeromodes}
in a volume $V= L^{3}\times T$ take the form \cite{Stani}
\bea
D_{| \nu |}^{(1)} &=& \frac{2 | \nu |}{(F_{\pi}L)^{2}} 
\left\{ | \nu | + \frac{\alpha}{2 N_{c}} - 
\frac{\beta_{1}}{F_{\pi}^{2} \sqrt{V}} \nn \right. \\
&& \left. + \Big[ \frac{\gamma_{1}}{2} - \frac{1}{24} \Big(
\frac{7}{3} + 2 \nu^{2} - 2 \langle \nu^{2} \rangle
\Big) + \frac{\gamma_{1}}{2} \Big] \,
\frac{T^{2}}{F_{\pi}^{2}V} \right\} \ , \\
D_{| \nu |}^{(2)} &=& - \frac{2 | \nu |}{(F_{\pi}L)^{2}} 
\left\{ 1 + | \nu | \Big( \frac{\alpha}{2 N_{c}} - 
\frac{\beta_{1}}{F_{\pi}^{2} \sqrt{V}} \Big) \right. \nn \\
&& \left. + | \nu | \Big[ \frac{\gamma_{1}}{2} - \frac{1}{24} 
\Big( \frac{13}{3} - 2 \langle \nu^{2} \rangle
\Big) \Big] \,
\frac{T^{2}}{F_{\pi}^{2}V} \right\} \ , \\
&& {\rm where~in ~our ~ case} \nn \\
&& \beta_{1} = 0.1314565 \ \ , \quad 
\gamma_{1} = - \frac{1}{12} \sum_{\vec p \neq \vec 0}
\frac{1}{\sinh ^{2} ( T | \vec p \, | /2)} = -0.083291 \ . \nn
\eea
$\beta_{1}$ is a shape coefficient, which we computed for our anisotropic
volume according to the prescription in Ref.\ \cite{epsFP}.

\begin{figure}[ht!]
  \centering
\includegraphics[angle=270,width=.8\linewidth]{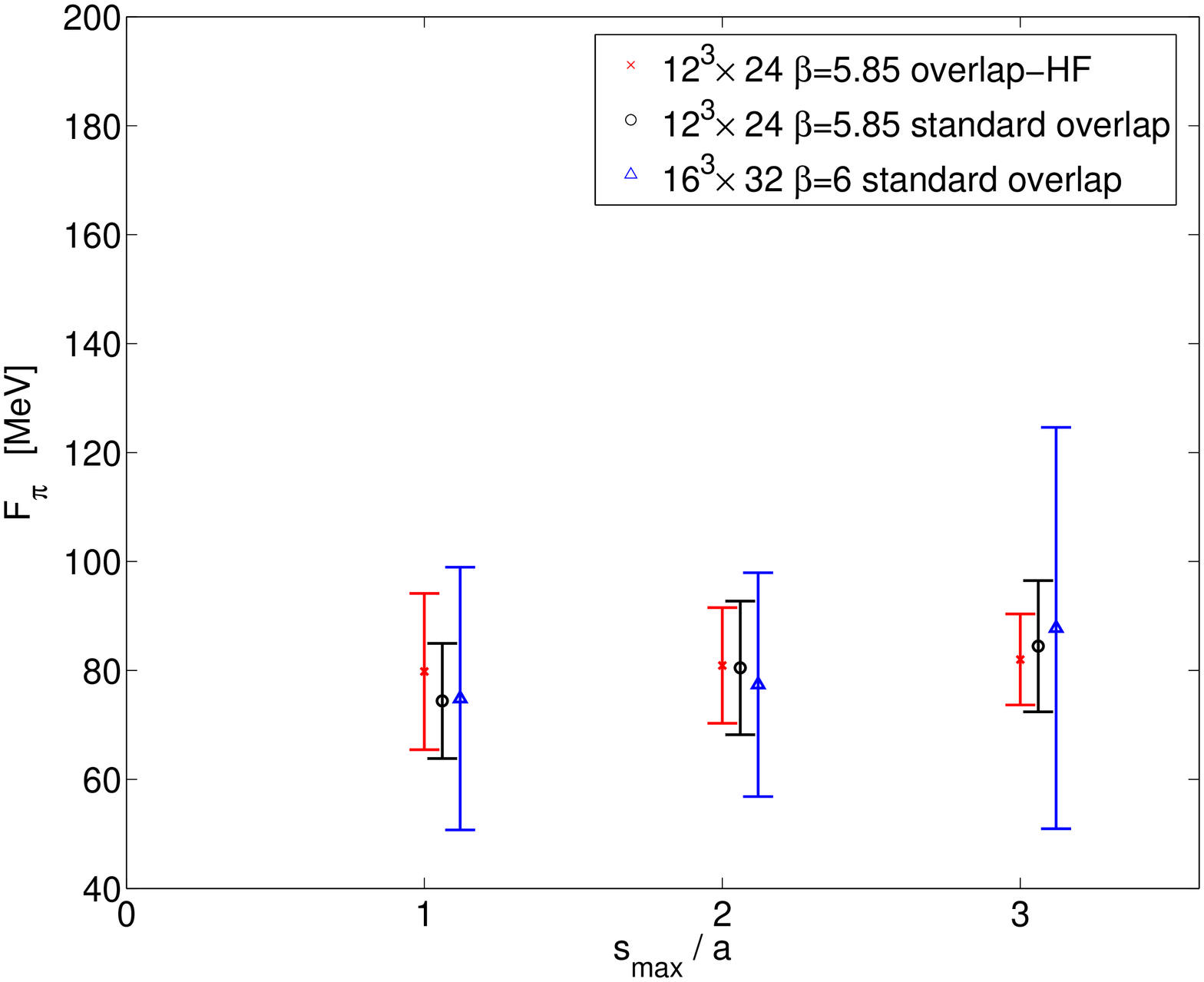}
\caption{{\it The results for $F_{\pi}$ based on a global fit
of our data to a quenched $\chi$PT prediction for the
zero-mode contributions to the pseudoscalar correlations
function. Here and in Fig.\ \ref{alphafig}
we show the results of a two parameter fit over the ranges
$s \in [T/2-s_{\rm max},T/2 + s_{\rm max}]$.}}
\label{Fpizerofig}
\end{figure}

\begin{figure}[h!]
  \centering
\includegraphics[angle=270,width=.8\linewidth]{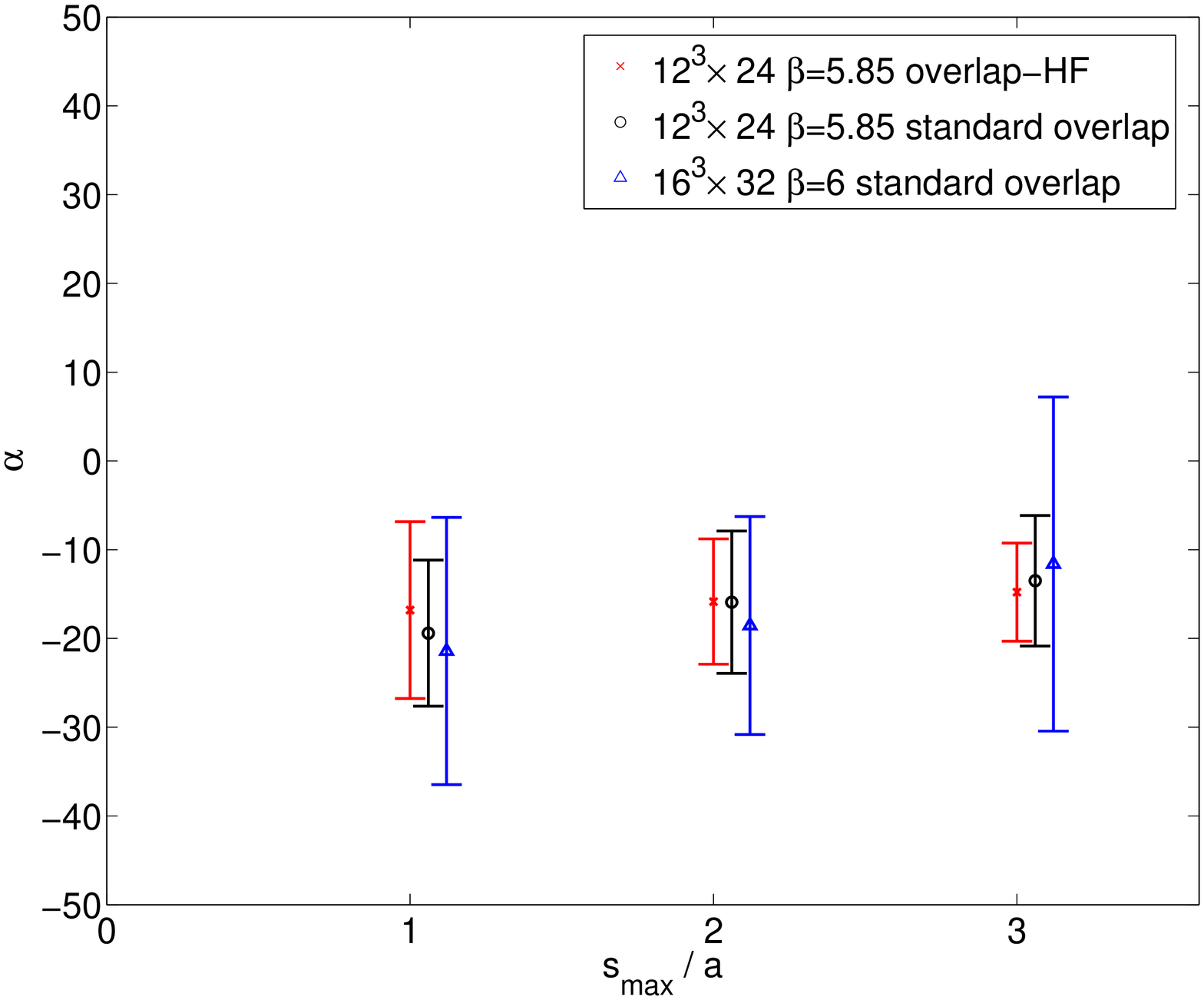}
\caption{{\it The results for the quenching specific LEC
$\alpha$, based on a global fit
of our data to a quenched $\chi$PT prediction for the
zero-mode contributions to the pseudoscalar correlations
function. Here and in Fig.\ \ref{Fpizerofig}
we show the results of a two parameter fit over the ranges
$s \in [T/2-s_{\rm max},T/2 + s_{\rm max}]$.}}
\label{alphafig}
\end{figure}

We evaluated the LECs $F_{\pi}$ and $\alpha$ from fits to the
linear term in eq.\ (\ref{Taylor}). We used all the zero-modes
that we identified in the topological sectors $| \nu |=1$ and $2$ ---
the statistics is given in Table \ref{epstab}.
For $\langle \nu^{2} \rangle$ (which enters the expressions for
$D_{|\nu |}^{(i)}$ through Witten-Veneziano relations) we inserted the
result that we measured in each case, which is also given in 
Table \ref{epstab}.
For each of our lattice sizes and each type of overlap operator
we performed a global fit over both topological sectors involved, 
in a fitting range $s_{\rm max}$.
The emerging optimal values for $F_{\pi}$ and $\alpha$ are shown
in Figs.\ \ref{Fpizerofig} and \ref{alphafig}, and the values
at $s_{\rm max}/a = 1$ are given in Table \ref{epstabzero}. 
We see that the results for different lattice spacings
and overlap Dirac operators are in good agreement, and we obtain
the most stable plateau for $D_{\rm ovHF}$. 

\begin{table}
\begin{center}
\begin{tabular}{|c||c|c|c|}
\hline
Dirac operator & $D_{\rm ovHF}$ & $D_{\rm N}$ & $D_{\rm N}$ \\
\hline
$\beta$ & $5.85$ & $5.85$ & $6$ \\
\hline
lattice size & $12^{3}\times 24$ & $12^{3}\times 24$ & $16^{3}\times 32$ \\
\hline
\hline
$F_{\pi}$ & $(80 \pm 14)$ MeV & $(74 \pm 11)$ MeV & $(75 \pm 24)$ MeV \\
\hline
$\alpha$  & $-17 \pm 10$ & $-19 \pm 8$ & $-21 \pm 15$ \\
\hline
\end{tabular}
\end{center}
\caption{{\it Our results in the $\eps$-regime for the pion
decay constant $F_{\pi}$ ---
along with the quenching specific LEC $\alpha$ ---
based on the zero-mode contributions to the
pseudoscalar correlation function, see Subsection 4.4. 
The joint statistics in the sectors $|\nu | =1$ and $2$,
given in Table \ref{epstab}, contributes. We give the results
at fitting range $s_{\rm max}=a$, which
is most adequate in view of eq.\ (\ref{Taylor}).}}
\label{epstabzero}
\end{table}

The value that we now obtain for $F_{\pi}$ is below
the one of Section 4.4, which used a different observable and
a different $\epsilon$-counting rule for the quenched terms.
In fact, the result of this Section is close to the phenomenological value
(we repeat that the latter amounts 
to $\approx 86 ~ {\rm MeV}$ if one extrapolates to
the chiral limit \cite{CoDu}).
This result, as well as the negative value for $\alpha$, are also somewhat
below the values reported in Ref.\ \cite{zeromodes} based on the same
method. Some differences are that Ref.\ \cite{zeromodes} always used
$D_{\rm N}$ (with various values of $\mu$), cubic volumes,
a continuum extrapolated value for $\langle \nu^{2} \rangle$
and partial fits were performed. We suspect that the anisotropic
shape of our volumes, $T = 2L$, could be the main source of the deviation
from those results \cite{Mikko}.

\section{Conclusions}

We have constructed an overlap hypercube Dirac operator $D_{\rm ovHF}$, 
which is especially suitable at a lattice spacing of
$a \simeq 0.123 ~{\rm fm}$. It has a strongly improved locality 
compared to the standard overlap operator $D_{\rm N}$.
This operator defines chiral fermions on coarser lattices than
$D_{\rm N}$.

We performed quenched simulations with $D_{\rm ovHF}$ and with
$D_{\rm N}$ in a volume 
$V \simeq (1.48 ~ {\rm fm})^{3} \times (2.96 ~ {\rm fm})$
at $\beta = 5.85$ and at $\beta =6$. \\

In the $p$-regime we applied $D_{\rm ovHF}$ and measured
the meson masses $m_{\pi}$ and $m_{\rho}$, the PCAC quark mass
$m_{\rm PCAC}$ and the pion decay constant $F_{\pi}$ at bare quark masses
ranging from $16.1 ~ {\rm MeV}$ to $161 ~ {\rm MeV}$. The results
for $m_{\pi}$ and $m_{\rho}$ are similar to the values
found previously with $D_{\rm N}$ on the same lattice, which
confirms their validity. On the other hand,
$m_{\rm PCAC}$ turned out to be much closer to $m_{q}$ than in the
standard overlap formulation. This implies an axial-current renormalisation
constant close to $1$, $Z_{A} = 1.17(2)$, which is favourable for the
connection to perturbation theory.
Regarding $F_{\pi}$, it turned out that the data obtained in the
$p$-regime can hardly be extrapolated to the chiral limit. \\

We considered a large number of topological charges defined by the
fermion indices of $D_{\rm ovHF}$ or of $D_{\rm N}$
(which coincide in part), and we
found histograms which approximate well a Gaussian.
The resulting topological susceptibility is in good agreement
with the literature.

In the $\epsilon$-regime we determined a value for the chiral condensate
from the distribution of the lowest eigenvalues. For both, 
$D_{\rm ovHF}$ and $D_{\rm N}$ we obtained $\Sigma$ values around
$(300 ~ {\rm MeV})^{3}$. 

We evaluated $F_{\pi}$ in the $\epsilon$-regime in two ways, from the 
axial-current correlation and from the zero-mode contributions to the 
correlation of the pseudoscalar density. These two methods handle the
$\epsilon$-counting of the quenched terms differently, and they yield
different values for $F_{\pi}$. The axial-current method leads
to $F_{\pi} \approx 110 ~{\rm MeV}$, which is consistent
with various other quenched results in the literature.
The zero-mode method 
(which might be more sensitive to the anisotropy of the volume)
leads to a lower $F_{\pi}$ around  $84 ~{\rm MeV}$, in the
vicinity of the phenomenological value. The final result of Ref.\ 
\cite{LMA} --- using again a different method, based on
the $\Delta I = 1/2$ rule,
still in the $\epsilon$-regime --- is in between.
We add that recently further methods were proposed to
evaluate $F_{\pi}$ in the $\eps$-regime,
involving $\Delta s =1$ transitions \cite{Ds1} and a chemical
potential \cite{chempot}. \\

From the current results, we conclude that the methods applied here work
in the sense that they do have the potential to evaluate at 
least the leading LECs from lattice simulations in the $\epsilon$-regime.
The quenched data match
the analytical predictions qualitatively (if the volume is not too small)
and --- in the setting we considered ---
they lead to results in the magnitude of the LECs in Nature.
However, the quenched results are ambiguous: different methods
yield different values.

For precise values and a detailed comparison 
to phenomenology, simulations with
dynamical quarks will be needed. In particular the $\epsilon$-regime
requires then dynamical Ginsparg-Wilson fermions. 
For instance, quenched re-weighting already leads to a distribution
of the microscopic Dirac eigenvalues and the topological charges
as it is expected for one dynamical quark flavour \cite{rewei}. 
In view of truly dynamical QCD simulations, first tests show
that it is hard to arrange for topological 
transitions \cite{dynov},
but fortunately the $\epsilon$-regime investigations can work even
in a fixed topological sector. Therefore, and also in view of the lattice
size, the $\epsilon$-regime is promising, if one is able to handle 
sufficiently small quark masses in a Hybrid Monte Carlo simulation,
and if ergodicity inside a fixed topological sector is achieved.\\

\noindent
{\bf Acknowledgements:} \
{\it We are indebted to M.\ Papinutto and C.\ Urbach for
numerical tools. We also thank A.\ Ali Khan,
S. D\"{u}rr, H.\ Fukaya, P.\ Hasenfratz, S.\ Hashimoto,
E.\ Laermann, M.\ Laine, K.-I.\ Nagai, K.\ Ogawa, 
L.\ Scorzato, A.\ Shindler, H.\ St\"{u}ben, P.\ Watson,
and U.\ Wenger for useful comments. 
This work was supported by the Deutsche Forschungsgemeinschaft 
through SFB/TR9-03. 
The computations were performed on the IBM p690 clusters of the
``Norddeutscher Verbund f\"ur Hoch- und H\"ochstleist\-ungs\-rechnen'' (HLRN) 
and at NIC, Forsch\-ungs\-zentrum J\"{u}lich.}

\end{document}